\documentclass[%
twocolumn,
superscriptaddress,
longbibliography,
nofootinbib,
 amsmath,amssymb,
 aps,
prb,
]{revtex4-2}

\usepackage{braket}
\usepackage{physics}
\usepackage{comment}
\usepackage{xcolor}
\usepackage{appendix}
\usepackage{caption}
\usepackage{subcaption}
\usepackage{graphicx}
\usepackage{dcolumn}
\usepackage[mathscr]{euscript}
\usepackage[scr=boondox]{mathalpha}
\usepackage{bm}
\usepackage[linktocpage]{hyperref}
\hypersetup{colorlinks=true,citecolor=blue,linkcolor=blue, urlcolor=blue, breaklinks=true}

\makeatletter
\newcommand\xlabel[2][]{\phantomsection\def\@currentlabelname{#1}\label{#2}}
\makeatother

\newcommand{\Z}{\mathbb{Z}}
\newcommand{\HH}{\mathcal{H}}

\begin{document}

\preprint{APS/123-QED}

\title{Fractional disclination charge and discrete shift in the Hofstadter butterfly}

\author{Yuxuan Zhang}
\affiliation{Department of Physics and Joint Quantum Institute, University of Maryland,
College Park, Maryland 20742, USA}
\affiliation{Condensed Matter Theory Center, University of Maryland,
College Park, Maryland 20742, USA}
\author{Naren Manjunath}
\affiliation{Department of Physics and Joint Quantum Institute, University of Maryland,
College Park, Maryland 20742, USA}
\affiliation{Condensed Matter Theory Center, University of Maryland,
College Park, Maryland 20742, USA}
\author{Gautam Nambiar}
\affiliation{Department of Physics and Joint Quantum Institute, University of Maryland,
College Park, Maryland 20742, USA}
\author{Maissam Barkeshli}
\affiliation{Department of Physics and Joint Quantum Institute, University of Maryland,
College Park, Maryland 20742, USA}
\affiliation{Condensed Matter Theory Center, University of Maryland,
College Park, Maryland 20742, USA}

\begin{abstract}
In the presence of crystalline symmetries, topological phases of matter acquire a host of invariants leading to non-trivial quantized responses. Here we study a particular invariant, the discrete shift $\mathscr{S}$, for the square lattice Hofstadter model of free fermions. $\mathscr{S}$ is associated with a $\Z_M$ classification in the presence of $M$-fold rotational symmetry and charge conservation. $\mathscr{S}$ gives quantized contributions to (i) the fractional charge bound to a lattice disclination, and (ii) the angular momentum of the ground state with an additional, symmetrically inserted magnetic flux. $\mathscr{S}$ forms its own `Hofstadter butterfly', which we numerically compute, refining the usual phase diagram of the Hofstadter model. We propose an empirical formula for $\mathscr{S}$ in terms of density and flux per plaquette for the Hofstadter bands, and we derive a number of general constraints. We show that bands with the same Chern number may have different values of $\mathscr{S}$, although odd and even Chern number bands always have half-integer and integer values of $\mathscr{S}$ respectively. 
\end{abstract}

\maketitle

Topological phases of matter are characterized by invariants that give rise to quantized physical responses, such as the Chern number and associated quantized Hall conductivity. In the presence of spatial symmetries, additional invariants also arise, such as the Wen-Zee shift \cite{Wen1992shift} in clean isotropic continuum quantum Hall systems, which characterizes the response to geometric curvature \cite{Avron1995hvisc,Read2009hvisc,Read2011hvisc,haldane2009hall,haldane2011fqh,Gromov2014,Bradlyn2015,Gromov2015,schine2016,wu2017fqh}.
In order to fully understand systems where lattice effects play an important role, we must develop a complete understanding of invariants and their associated quantized responses for topological phases with charge conservation and crystalline symmetries. Recently, \cite{manjunath2021cgt,Manjunath2020fqh} developed such a systematic theory in two dimensions by analyzing topological terms for crystalline gauge fields and found several additional quantized invariants. Some of these invariants have no continuum analog, while others give a crystalline analog of invariants known from the setting of continuum spatial symmetries. 

In this paper, we study one such invariant, the discrete shift $\mathscr{S}$, and its physical consequences through numerical studies of the square lattice Hofstadter model \cite{hofstadter1976} of spinless free fermions. $\mathscr{S}$ is an invariant that depends on a discrete $\Z_M$ rotational symmetry and $U(1)$ charge conservation, and is a discrete analog of the Wen-Zee shift arising in continuum systems. For invertible fermionic topological states \cite{barkeshli2021invertible,freed2016,aasen2021characterization}, $2\mathscr{S}$ is an integer defined mod $2M$;  we show that for fixed Chern number, $\mathscr{S}$ can in principle take one of $M$ distinct values, and odd (even) Chern numbers must have half-integer (integer) values of $\mathscr{S}$.\footnote{For invertible bosonic topological states, $\mathscr{S}$ must be integer, while for fractionalized topological states, $\mathscr{S}$ can be fractional. } 

\begin{figure}[t]
\includegraphics[width=0.43\textwidth]{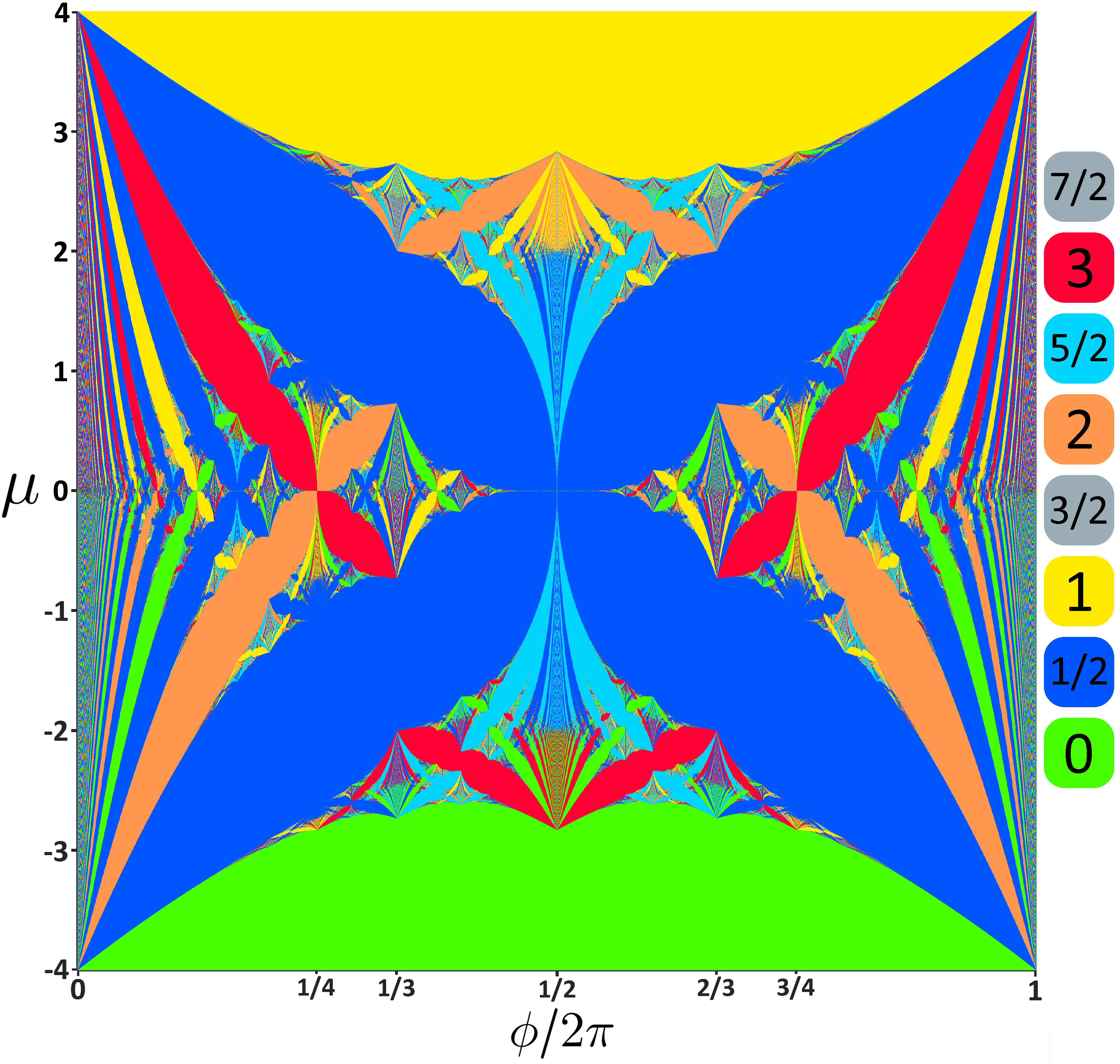}
\caption{\label{fig:shift} $\mathscr{S}$ for Hofstadter model, from Eq.~\eqref{eq:s_formula}.}
\end{figure}

Remarkably, $\mathscr{S}$ refines the known phase diagram of the Hofstadter model, leading to a new Hofstadter butterfly (Fig.~\ref{fig:shift}), which we numerically compute. As we study numerically in detail, $\mathscr{S}$ has a physical manifestation in terms of a quantized contribution to the fractional charge bound to lattice disclinations (see Eq.~\eqref{eq:QWfield}) and, dually, the fractional angular momentum bound to magnetic flux (see Eq.\eqref{eq:l}). We theoretically justify several properties of $\mathscr{S}$ that are evident from Fig. \ref{fig:shift}, and also propose an empirical formula for $\mathscr{S}$ (Eq.~\eqref{eq:s_formula}). 

Since the Hofstadter model has now been effectively realized in moir{\'e} superlattice systems \cite{Dean2013,Hunt2013hb,Saito2021,Eric2018moire}, ultracold atoms \cite{aidelsburger2013,miyake2013,kennedy2015}, and photonics \cite{hafezi2013,ozawa2019}, an experimental verification of our results may be possible.

We note that \cite{Biswas2016,Liu2019ShiftIns,Li2020disc,You2020hoe,2108.00008,peterson2021trapped,Han2019} also study some aspects of the shift in lattice settings, with limited results when Chern number $C \neq 0$, as discussed below and in Appendix \ref{sec:priorwork}. 

\paragraph*{Model and response theory.} 
We consider a system of fermions hopping on a lattice with a discrete $\Z_4$ rotational symmetry, a background magnetic flux $\phi$ per unit cell, and filling $\nu_0$ charge per unit cell. We focus on the Hofstadter model on a square lattice, with the Hamiltonian
$H=-t\sum_{\langle i,j \rangle }c_{i}^{\dagger}c_{j}e^{-iA_{bgd;ij}}+h.c$.
This describes spinless free fermions $c_i$ coupled to a background $U(1)$ gauge field $A_{bgd}$, whose holonomy around each plaquette is $\phi$. 
When $\phi = 2\pi\frac{p}{q}$ with coprime integers $p,q$, the system has $q$ bands. When $r$ bands are filled, $\nu_0 = r/q$.
At any gapped point in the parameter space $(\nu_0,\phi)$, the total Chern number $C$ of the filled bands is determined by the conditions $\nu_0 = \frac{p}{q} C \mod 1; \quad |C| \le \frac{q}{2}$ \cite{thouless1982}. 
Each connected gapped region in this parameter space has a fixed value of $C$. Simply connected gapped regions with Chern number $C$ (referred to as Chern number $C$ lobes) are separated by special values of $\frac{\phi}{2\pi}$ which lie in the Farey sequence of order $2 |C|$ \cite{osadchy2001db}. The continuum limit of $n$ filled Landau levels is obtained by taking $\frac{p}{q} \rightarrow 0^+, C=n$. 

$H$ has a symmetry group $G$ which is a central extension of the wallpaper group p4 = $\Z^2 \rtimes \Z_4$ by $U(1)$. This means that the magnetic translations are generated by the many-body operators $T_{\mathbf{x}},T_{\mathbf{y}}$, which satisfy
  $T_{\mathbf{x}}T_{\mathbf{y}} = T_{\mathbf{y}}T_{\mathbf{x}} e^{i \phi \hat{N}}$,  
where $\hat{N}$ is the total particle number operator. The Hamiltonian is invariant under a ``magnetic" rotation operator $\tilde{C}_{4,\lambda} \equiv \hat{C_4}e^{i\sum_j \lambda_{j}c_j^{\dagger}c_j}$ where $\lambda_j$ is a gauge transformation at site $j$ which is fixed up to an overall constant by $A_{bgd}$. The usual rotation operator $\hat{C}_4$ acts as $\hat{C}_4 c_j \hat{C}_4^\dagger = c_{R(j)}$, where $R(j)$ is a vertex-centered $\pi/2$ rotation of site $j$.


The quantized universal properties can be encoded by a topological response theory involving a background $G$ gauge field $B = (\delta A, \vec{R},\omega)$, which is nonabelian \cite{manjunath2021cgt}. Here $\delta A = A - A_{bgd}$ is the deviation of the total $U(1)$ gauge field $A$ relative to $A_{bgd}$, while $\omega$, $\vec{R}$ are the crystalline gauge fields (Appendix \ref{sec:CGTreview} gives additional details). $\omega$ is a background gauge field for the $\Z_4$ rotational symmetry; in the continuum limit, it is identified with the spin connection. $\vec{R}$ is a two-component gauge field for the $\Z^2$ translational symmetry. In terms of $\vec{R}$ and $\omega$ we also define an area element $A_{XY}$ which counts the number of unit cells \cite{manjunath2021cgt}. The integral $\delta \Phi_W \equiv \int_W d\delta A$ over a two-dimensional region $W$ gives the total excess magnetic flux within $W$, not including the background flux, while $\int_W d\omega$ gives the total disclination angle  of disclinations within $W$. $\omega$, $\vec{R}$ are taken to be real-valued fields with quantized periods, since they are $\Z_4$ and $\Z^2$ gauge fields respectively.\footnote{One can also work in a simplicial formulation where $\omega$, $\vec{R}$ are taken to have discrete values \cite{manjunath2021cgt}.} 

We can write all terms in the topological response theory which depend on $A$ or $\omega$ \cite{manjunath2021cgt} (note that the response theory involves $A$ and not just $\delta A$):
\begin{align}\label{eq:chargeresponse}
  \mathcal{L} &=  \frac{C}{4\pi} A \wedge dA + \frac{\mathscr{S}}{2\pi} A \wedge d\omega + \frac{\vec{\mathscr{P}}_c}{2\pi} \cdot A \wedge \vec{T} + \frac{k_0}{2\pi}A \wedge A_{XY} \nonumber \\
   &+ \frac{\tilde{\ell}_s}{4\pi} \omega \wedge d\omega + \frac{\vec{\mathscr{P}}_s}{2\pi} \cdot \omega \wedge \vec{T} +\frac{k_s}{2\pi} \omega \wedge A_{XY} .
\end{align}
See Appendix \ref{sec:CGTreview} for a discussion. $\vec{T} = d\vec{R} + i \sigma_y \omega \wedge \vec{R}$ is the torsion 2-form. Here $C,2\mathscr{S}$ must be quantized to integers. For fermionic invertible phases there are some additional terms in the theory, discussed in Appendix \ref{sec:CGTreview}. The first term defines the Hall conductivity $\overline{\sigma}_H = \frac{C}{2\pi}$ in natural units and assigns charge to flux. The second term assigns a fractional $U(1)$ charge $\mathscr{S}\frac{\Omega}{2\pi}$ to a defect with disclination angle  $\Omega$. On a square lattice, it is topologically trivial if a $\frac{\pi}{2}$ disclination is assigned an integer charge, which can be removed by applying local operators at the disclination core. Thus only $\mathscr{S} \mod 4$ is a symmetry-protected invariant, in contrast to the continuum shift, which is a $\Z$ invariant.\footnote{Note that the conventional definition of shift in the quantum Hall literature is $S \equiv \chi \frac{\mathscr{S}}{\overline\sigma_H}$, with $\chi$ the Euler characteristic of the space.} The second term can also be written as $\frac{\mathscr{S}}{2\pi} \omega \wedge d A$, which assigns angular momentum $\mathscr{S} \frac{dA}{2\pi}$ to flux $dA$. In Appendix~\ref{Sec:properties_S} we show the nontrivial result that for spinless fermions in free or interacting systems, $\mathscr{S}$ is quantized to a half-integer if $C$ is odd, and to integers if $C$ is even. The numerical values of $\mathscr{S}$ in Fig. \ref{fig:shift} agree with this result.

$\nu_0 = C \frac{\phi}{2\pi} + k_0$ and $\nu_s = \mathscr{S} \frac{\phi}{2\pi} + k_s$ are the charge and angular momentum per unit cell, with $k_0$, $k_s$ integers. The terms with $\vec{T}$ in Eq.~\eqref{eq:chargeresponse} can be detected by inserting defects with nontrivial dislocation Burgers vector, but we do not consider such defects in this work. Hereafter we ignore these, as well as the $\omega \wedge d\omega$ term.

\paragraph*{Fractional disclination charge.}

Eq.~\eqref{eq:chargeresponse} predicts a contribution to the charge localized at a $\frac{\pi}{2}$ disclination coming from $\mathscr{S}$. Here we compare the field theory prediction to microscopic calculations. The discussion below applies to general lattices with p4 space group symmetry.  

We construct a $\frac{\pi}{2}$ disclination at the point $o$ by deleting a quadrant from the infinite plane and reconnecting sites using the operator $\tilde{C}_{4,\lambda}$ (see Appendix \ref{sec:Hdisc}). In particular, if each unit cell in the disclination lattice has the same flux, we show that $\lambda_o = 0$. Now consider a region $W$ enclosing the disclination, such that the distance between the disclination and the boundary $\partial W$ is much greater than the correlation length. The total charge is 
\begin{equation}
\label{eq:QWfield}
    Q_W = \int_W \frac{\delta \mathcal{L}}{\delta A_0} = C \frac{\delta \Phi_W}{2\pi} + \mathscr{S} \frac{\Omega_W}{2\pi} + \nu_0 n_{u.c.,W},
\end{equation}
where $n_{u.c.,W}$, $\Omega_W$ and $\delta \Phi_W$, are the number of unit cells in $W$, disclination angle, and excess magnetic flux (on top of the background flux $n_{u.c.,W}\phi$) respectively. Here we use that $\int_W d A = \delta \Phi_W + n_{u.c.,W} \phi$ and $\nu_0 = C \phi/2\pi + k_0$. In order to use Eq.~\eqref{eq:QWfield}, $W$ should enclose a definite integer number of unit cells. Furthermore, in order to ensure that the condition $Q_W + Q_{W'} = Q_{W \cup W'}$ holds microscopically, as in the field theory, we define $Q_W \equiv \sum_{i \in W} \text{wt}(i) Q_i$ where the weight $\text{wt}(i) = 1$ for interior points and $\text{wt}(i)=\frac{1}{4},\frac{2}{4},\frac{3}{4}$ if the interior of $W$ subtends an angle $\frac{\pi}{2}, \pi, \frac{3\pi}{2}$ at site $i$ (see Fig. \nameref{fig:2A}). Note our definition of disclination charge differs from previous work. \cite{Li2020disc}

\begin{figure}[t]
    \centering
    \includegraphics[width=0.485\textwidth]{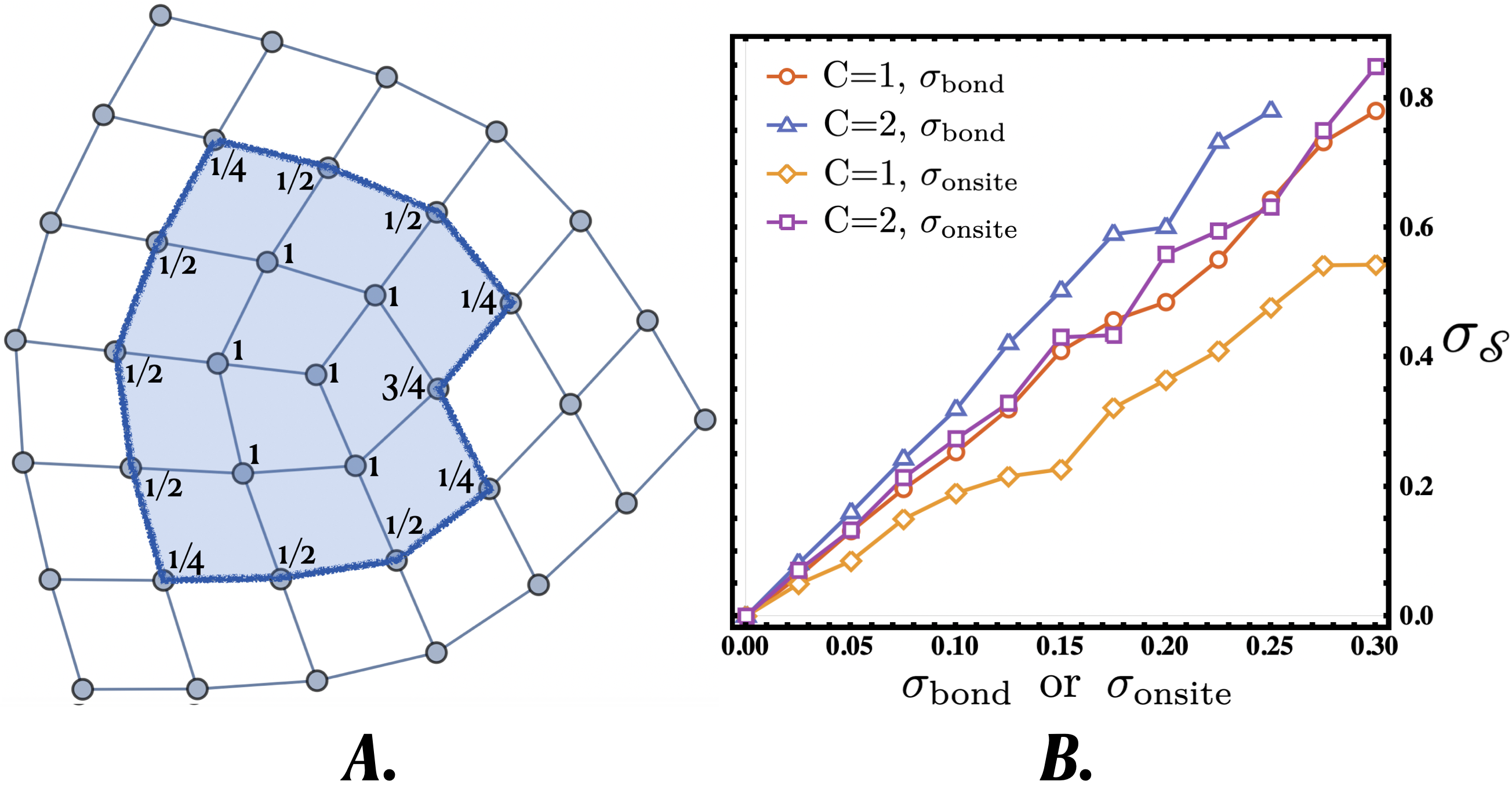}
	
    \caption{\textbf{A.} Lattice disclination with disclination angle $\Omega=\frac{\pi}{2}$. The blue region $W$ covers 11 unit cells. $Q_i$ is weighted by the indicated amount when calculating $Q_W$ or $\bar{Q}_W$ \xlabel[2A]{fig:2A}. \textbf{B.} Standard deviation of $\mathscr{S}$ as a function of bond disorder $\sigma_{\text{bond}}$ or onsite disorder $\sigma_{\text{onsite}}$ for $C=1$ and $C=2$ main Landau level (average hopping is 1).\xlabel[2B]{fig:2B}
    }
    
\end{figure}

\begin{figure}[t]
\includegraphics[width=0.46\textwidth]{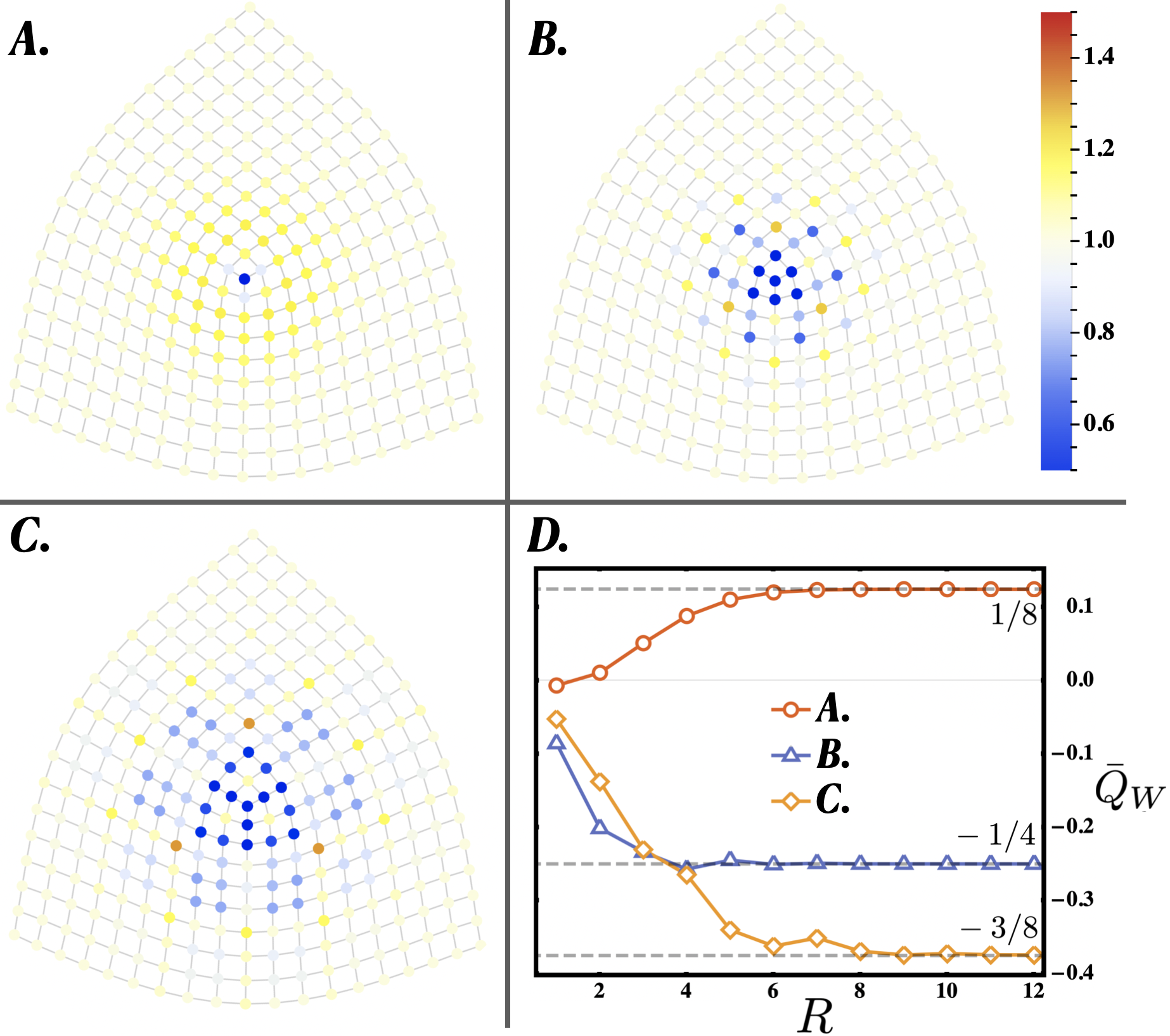}
\caption{\label{fig:Wannier} \textbf{(A-C):} $Q_i$ for each site $i$ in a clean system with parameters \textbf{A.} $\frac{\phi}{2\pi}=\epsilon, C=1$; \textbf{B.} $\frac{\phi}{2\pi}=\frac{1}{2}-\epsilon$, $C=-2$; \textbf{C.} $\frac{\phi}{2\pi}=\frac{1}{3}+\epsilon, C=3$. The colorbar is in units of $\nu_0$. $\epsilon$ is a small fraction which opens up the band gap. \textbf{D.} $\bar{Q}_W$ for \textbf{(A-C)} is quantized at $\frac{\mathscr{S}}{4}$ 
when $R\ge 9$. $R$ is the distance from the disclination center to $\partial W$. (total side length is $L=24$ unit cells). Here we have cropped out the edges to show the bulk features more clearly; the full figures with edges are given Appendix .\ref{Sec:disc_sublattice}}
\end{figure}

The charge on a site $Q_i$ is simply $\expval{c_i^\dagger c_i}$ in the ground state. We choose $W$ such that it encloses a single disclination with $\Omega_W=\frac{\pi}{2}$. We set the excess flux $\delta \Phi_W =0$ in our numerics. $\nu_0$ can be defined as the filling of a corresponding clean lattice on a torus with the same flux per unit cell $\phi$. On a lattice with disclinations, $\nu_0$ is also the charge per unit cell far away from the disclination. Suppose we define the excess charge in $W$ as $\bar{Q}_W\equiv Q_W - \nu_0 n_{u.c.,W}$. We can then extract the shift $\mathscr{S}$ to be $\frac{\mathscr{S}}{4}=\bar{Q}_W$. Numerically, we find that $\bar{Q}_{W}$, and hence the computed $\mathscr{S}$ is indeed independent of the size of $W$ for large enough $W$. We show this by explicitly plotting $Q_i$ and the size dependence of $\bar{Q}_W$ for three representative sets of parameters in the Hofstadter butterfly (Fig.~\ref{fig:Wannier}). We can thus use this procedure to calculate $\mathscr{S}$ throughout the Hofstadter butterfly; this is shown in Fig.~\ref{fig:shift}. In the Landau level limit, we numerically recover the result $\mathscr{S}=\frac{C^2}{2}$ \cite{Wen1992shift}.

It is instructive to apply Eq.~\eqref{eq:QWfield} when $W$ is the entire surface of a cube, which has 8 $\frac{\pi}{2}$ disclinations corresponding to the 8 corners. In this case, we obtain 
\begin{equation}\label{eq:Qcube}
    Q_{\text{cube}}= 2\mathscr{S}+\nu_0 n_{u.c.,\text{cube}}.
\end{equation}
Thus, in order to be in the same gapped phase as a state on a torus with filling $\nu_0$ and identical $\phi$ and Chern number $C$, the total charge on the cube is shifted from the na\"ive expectation of $\nu_0 n_{u.c., \text{cube}}$ by $2\mathscr{S}$. Numerically, this agrees with the number of extra single particle states we need to fill. 

Note that the weighting procedure and the numerical result $\bar{Q}_W = \frac{\mathscr{S}}{4}$ generalize to any $C_4$ symmetric lattice. The details are described in Appendix \ref{Sec:disc_sublattice}.

Let us now introduce bond and on-site potential disorder, which break the crystalline symmetry. In this case, the value of $\mathscr{S}$ extracted from $\bar{Q}_W$ through $\mathscr{S} = 4 \bar{Q}_W$ deviates from its quantized value for each disorder realization, although remains quantized after disorder averaging. The standard deviation $\sigma_{\mathscr{S}}$ computed from $\bar{Q}_{W}$ grows to order 1 with an increase in disorder strength, as shown for two representative lobes in Fig.~ \nameref{fig:2B}. The value of $\mathscr{S}$ extracted from $Q_{\text{cube}}$ is much more robust (i.e. much smaller standard deviation) as long as the chemical potential is far from the band edge, since in this case $Q_{\text{cube}}$ can change only if a given disorder configuration moves a single particle state across the chemical potential.

\paragraph*{Angular momentum due to flux}\label{se:clean}
Since $\omega$ is a rotation gauge field, the angular momentum is the charge under rotations, given by $l = \int \frac{\delta \mathcal{L}}{\delta \omega_0} = \mathscr{S} (\frac{\delta \Phi}{2\pi} + \frac{\phi}{2\pi} n_{u.c.}) + k_s n_{u.c.} \mod 4$. Below we compare this topological field theory (TFT) prediction to microscopic calculations. 
Consider the Hofstadter Hamiltonian on an $L \times L$ torus, with $L$ even and $\phi$ flux per plaquette. By turning on $\delta A$, we add a flux $2\pi \Delta m$, also distributed uniformly. If the system has Chern number $C$, we fill $\Delta m C$ additional single particle states in order to get a gapped many body state with the same $C$. This state has $m=\frac{\phi L^2}{2\pi}+\Delta m$ flux quanta.

We use the same magnetic rotation operator $\tilde{C}_{4,\lambda}$ that we used to define the disclination. This means that we require $\lambda_{o} = 0$, where $o$ is a fixed point of the rotation. 

Note that the holonomies $e^{-i \oint (\delta A+A_{bgd}) \cdot dl}$ along the two non-contractible cycles of the torus are position dependent. Translation by one site changes the holonomy by a factor $e^{i 2\pi \frac{m}{L}}$, and is therefore an exact symmetry only for the infinite system; for any finite system, translation by one site can only be an approximate symmetry. On a finite size torus with even $L$, there are two points, $o_1$ and $o_2$, distinguished by having holonomy $1$ or $e^{i \pi m}$ along both directions. The vertex-centered $\pi/2$ rotational symmetry is only exact for a finite size system when $o_1$ and $o_2$ are both fixed points of this rotation. 


Since we have two distinct fixed points $o_1$ and $o_2$, there are two distinct choices of gauge satisfying the above condition, either $\lambda_{o_1} = 0$ and $\lambda_{o_2} = m \frac{\pi}{2}$, or $\lambda_{o_2} = 0$ and $\lambda_{o_1} = - m \frac{\pi}{2}$. In what follows we pick the first choice, denoted $\tilde{C}_4$; the second is related by an overall $U(1)$ rotation $e^{- i \hat{N} m \frac{\pi}{2} }$, as will be discussed in Appendix \ref{sec:AMappendix}. 

The many-body ground state $|\Psi_m\rangle$ satisfies 
$\tilde{C}_{4}\ket{\Psi_m}=e^{il(m) \frac{\pi}{2}}\ket{\Psi_m}$.
Since $\tilde{C}_4^4 = 1$, $l(m)$ is an integer mod 4. 

In this setup, we require each rotation center to be a vertex. On a torus, this forces $L$ to be even: if $L$ were odd, any rotation would leave two points invariant, one at a vertex, and the other at a plaquette center. If the rotation center was at a plaquette center, then the original $\tilde{C}_4$ rotation would be modified by a lattice translation. The associated eigenvalue would receive a contribution from the $A \wedge \vec{T}$ term in Eq.~\eqref{eq:chargeresponse}, which we do not wish to consider here. Indeed, the numerical result with plaquette centred rotations is not consistent with Fig.~\ref{fig:shift}.

We find from direct numerical calculation:
\begin{equation}\label{eq:l}
    l(m)=\mathscr{S}m + C\frac{m^2}{2}+K(C,L) \mod 4.
\end{equation}
The quadratic dependence on $m$ is beyond the TFT description; while it is well-known from the continuum Landau level problem, it has not been derived using effective field theory.\footnote{\cite{Liu2019ShiftIns} accounted for the $m^2$ contribution in terms of the angular momentum of the electromagnetic field, however physically this term arises from the electron fluid itself.}
For a given lobe, $K(C,L)$ is an integer which does not depend on $m$, but does depend on $C$ and $L$. We can obtain $\mathscr{S}$ by subtracting the quadratic term and taking the difference: $\mathscr{S} = l(m+1)-l(m) - Cm - \frac{C}{2} \mod 4.$ For each lobe with a given $C$, the value of $\mathscr{S}$ obtained from Eq.~\eqref{eq:l} matches the result using Eq.~\eqref{eq:Qcube}, confirming the expected duality. 

Instead of inserting additional $\Delta m$ flux uniformly everywhere, we can insert it locally in a smaller region $W$ symmetrically around the rotation center. We find that the value of $\mathscr{S}$ remains constant for different sizes of $W$, even in the limit when $W$ contains just 4 plaquettes (when $C$ is small enough).

We can also extract $\mathscr{S}$ using \textit{partial} rotations \cite{You2020hoe,shiozaki2017invt}. That is, in a system with background flux $2\pi m_0$ we insert a local flux of $2\pi\Delta m$ in a region $W$, and we compute $\langle \Psi_{\Delta m, m_0} | \tilde{C}_{4}|_{D} |\Psi_{\Delta m, m_0}\rangle$, where $\tilde{C}_{4}|_{D}$ is the restriction of $\tilde{C}_{4}$ to a region $D$ containing $W$ and $|\Psi_{\Delta m, m_0}\rangle$ is the ground state. Let us fix $W$ and $D$ to be centered on $o_2$ and continue with our previous gauge choice $\lambda_{o_1} = 0$. 
The ground state expectation value can be written as $\bra{\Psi_{\Delta m, m_0}} \tilde{C}_4|_D\ket{\Psi_{\Delta m, m_0}}=e^{ - \gamma_D(\Delta m, m_0) + i l_{D}(\Delta m, m_0)\pi/2}$. The magnitude $e^{-\gamma_D(\Delta m, m_0)}$ has an exponentially decaying envelope as the perimeter $\partial D$ increases, as expected, in addition to an oscillatory behavior that we do not study in detail.
We empirically find 
$l_{D}(\Delta m, m_0) =  l_{D}(0,m_0) +\mathscr{S}\Delta m+ Cm\Delta m + \frac{C(\Delta m)^2}{2} \mod 4$. This matches the expectation from Eq.~\eqref{eq:l}. One can also perform the partial rotation computation in the case where the total system is defined with open boundary conditions, as discussed in Appendix \ref{sec:AMappendix}.

On the torus we find that the formula for $l_D(\Delta m, m_0) - l_D(0, m_0)$ is sensitive in complicated ways to the gauge choice $\lambda_{o_1} = 0$ or $\lambda_{o_2} = 0$, and whether $D$ is centered on $o_1$ or $o_2$; these dependencies are not fully understood. We discuss this and related issues arising for open boundary conditions in Appendix \ref{sec:AMappendix}.


\paragraph*{Theoretical analysis.}
As a function of $\mu$ and $\phi$, $\mathscr{S}$ and $C$ have the following general properties (note $\mathscr{S}$ is defined $\mod 4$):
\begin{enumerate}
    \item $\mathscr{S} \text{ mod } 1 = \frac{C}{2} \text{ mod 1}$.
    \item $\mathscr{S}(\mu,\phi) = \mathscr{S}(\mu,2\pi - \phi)$, i.e. $\mathscr{S}$ is invariant under time-reversal. 
    \item For the bands with the same Chern number $C$, $\mathscr{S}(\mu,\phi) = 1-\mathscr{S}(-\mu,\phi)$. 
    \item  When $\mathscr{S}$ changes,
$\frac{\phi}{2\pi}$ must lies in the Farey sequence of order $|C|$.
    \item $\mathscr{S}(\mu,0^+) = \frac{C^2}{2}$ for $C > 0$.
\end{enumerate}
Properties (1-3) will be justified in Appendix~\ref{Sec:properties_S}. We explain (1) also for general interacting systems, using the classification of invertible topological phases in Ref.~\cite{barkeshli2021invertible}. (2) follows for general interacting systems from the time-reversal invariance of the field theory term $\omega \wedge d A$, while (5) reproduces the known results in the continuum Landau level limit \cite{Wen1992shift}.

Now let us explain property (4). 
Consider all possible fractions $\frac{p}{q}$ with $0 \le \frac{p}{q} \le 1$, $1\le q \le 2|C|$ and $p,q$ coprime. Arrange them in increasing order, with 0 being the first element and 1 being the last. The resulting sequence is called the \textit{Farey sequence} of order $2|C|$. Now, the different lobes with Chern number $C$ are uniquely specified by the intervals of $\frac{\phi}{2\pi}$ obtained from this sequence \cite{osadchy2001db}. Moreover, each lobe has a constant value of shift. In our numerics,  when $\mathscr{S}$ jumps, $\phi/2\pi$ must lie at fractions $\frac{p}{q}$ in a smaller set, namely the Farey sequence of order $|C|$: see Appendix~\ref{sec:Farey}. 
\paragraph*{Empirical formula for $\mathscr{S}$.} Suppose we fix a $C > 0$ and consider $\mathscr{S(\phi)}$ as $\frac{\phi}{2\pi} = \frac{p}{q}$ is increased from 0 to 1. As stated above, $\mathscr{S}(\phi)$ jumps by integers at specific values of $\frac{p}{q}$, where $q \le |C|$. At a given $\frac{p}{q}$, the total jump, defined as $\lim\limits_{\epsilon \rightarrow 0^+} \mathscr{S}(2\pi \frac{p}{q} + \epsilon)-\mathscr{S}(2\pi \frac{p}{q}-\epsilon)$, is the sum of two contributions: 
A contribution of $-C-1$ whenever $q$ divides $C$, and another contribution of $2\lfloor \frac{C+q}{2q} \rfloor$ whenever $q$ is odd. 
Both contributions are automatically 0 if $q>|C|$. The observed jumps are tabulated in Appendix~\ref{sec:Farey} up to $C=12$. From these observations, we propose the following empirical formula by summing over all jumps that occur at $2\pi \frac{p}{q} <\phi$. For $C > 0$, 

\begin{equation}
\label{eq:s_formula}
    \mathscr{S}(\phi)= \frac{C^2}{2}  -(C+1)\left\lfloor\frac{C\phi}{2\pi}\right\rfloor
    + 2\sum_{\substack{\frac{p}{q}<\frac{\phi}{2\pi}\\\text{odd }q
    }} \left\lfloor \frac{C+q}{2q} \right\rfloor  \mod 4,
\end{equation}
where in the third term we sum over all $\frac{p}{q}$ in the Farey sequence of order $C$ that satisfy $\frac{p}{q}<\frac{\phi}{2\pi}$ and $q$ odd.
$\mathscr{S}$ for $C < 0$ can be obtained from the symmetry transformation $\mathscr{S}(\mu,\phi) = 1-\mathscr{S}(-\mu,\phi)$ which flips the sign of $C$.
We numerically checked Eq.~\eqref{eq:s_formula} for all $|C|\le 7$ lobes using Eq.~\eqref{eq:Qcube} on a cube of side length $L=28$. We also checked all $|C|\le 12$ lobes using Eq.~\eqref{eq:l} on a torus with side length $L=140$, and we checked some representative $|C|=45$ lobes with $L=180$.
We use \eqref{eq:s_formula}, together with an eigenvalue database \cite{osadchy2001db}, to generate Fig.~\ref{fig:shift}. 

Lobes with the same $(C, k_0)$ can have distinct values of $\mathscr{S}$. For example when $(C,k_0) = (4,-1)$ we can take $\frac{\phi}{2\pi} = 1/3 \mp \epsilon$, and find $\mathscr{S} = 3,1 \mod 4$ respectively. Also note that in Fig.~\ref{fig:shift}, we do not see any bands with total shift $\mathscr{S}=3/2$ or 7/2. However, there is no theoretical obstruction to realizing this in a system with odd $C$. Indeed, there are several examples of single excited bands that have odd $C$ and $\mathscr{S}=3/2$ or $7/2$ in this model. 

\paragraph*{Acknowledgements.} We thank M. Hafezi and S. Das Sarma for comments on the draft, and V. Galitski and D. Bulmash for discussions on related projects. This work is supported by the Laboratory for Physical Sciences through the Condensed Matter Theory Center, NSF CAREER (DMR- 1753240) (MB, NM), ARO W911NF-20-1-0232 (GN).
\bibliography{hofstadter_refs}

\appendix

\section{Relation to prior work}\label{sec:priorwork}

Here we summarize how our results relate to those in several prior works where the physical manifestation of the Wen-Zee shift is studied in a lattice setting.

Our contributions include:
\begin{enumerate}

\item A method to unambiguously extract the discrete shift by computing the charge $Q_W = \sum_{i \in W} \text{wt}(i) Q_i$ in a region $W$ containing a disclination, or through the angular momentum $l(m)$. Crucially, one must use the same rotation operator to define the disclination Hamiltonian and to compute the angular momentum. An important ingredient in our formula which was not introduced previously is the fractional weighting $\text{wt}(i) = 1/4,1/2,3/4$ for sites $i$ on the boundary $\partial W$. This fractional weighting is crucial to obtaining a simple formula that matches the topological field theory result. 

\item We have shown that the shift $\mathscr{S}$ is a half-integer defined modulo $M$ in the case of $M$-fold rotational symmetry, and that $\mathscr{S}$ is half-integer or integer depending on whether $C$ is odd or even, and we have derived these statements even in the case of interacting fermions. 

\item We have also provided an empirical formula for $\mathscr{S}$ everywhere in the Hofstadter butterfly for the square lattice Hofstadter model. 
\end{enumerate}

Ref.~\cite{Liu2019ShiftIns} discusses disclination charge in a 
model on the honeycomb lattice with $\mathbb{Z}_6$ rotational symmetry and particle-hole symmetry in the absence of a background magnetic field. They exploited the particle-hole symmetry to derive an analytical formula for disclination charge, and therefore the shift. 
We studied the disclination charge on the square lattice numerically in the presence of a background magnetic field (allowing for a tunable Chern number). In order to do so, we arrived at the prescription $Q_W = \sum_i \text{wt}(i) Q_i = \mathscr{S}/4 + \nu_0 n_{u.c.,W}$ that allows us to isolate the shift, given a very general charge distribution in a state.

Ref.~\cite{Liu2019ShiftIns} also studied the angular momentum response using global plaquette centered $2\pi/6$ rotations. We note that in our case, i.e., for 4-fold rotational symmetries, there exist both vertex- and plaquette-centred rotations, which are related to each other by a lattice translation; the latter may also introduce contributions from the field theory term $\vec{\mathscr{P}}_c \cdot A \wedge \vec{T}$. However, the model considered in Ref.~\cite{Liu2019ShiftIns} has only plaquette rotation centers, so this distinction does not arise in their case. The extra field theory term $\vec{\mathscr{P}}_c \cdot A \wedge \vec{T}$ is also expected to be trivial for systems with $6$-fold rotational symmetry \cite{manjunath2021cgt, Manjunath2020fqh}. We also note that gauge transformations required to accompany the rotation operator are nontrivial on the square lattice, because we cannot use a symmetric gauge (w.r.t. a vertex rotation center) to insert $2\pi$ flux. But this is not an issue in Ref.~\cite{Liu2019ShiftIns}, as a symmetric gauge can be used for $C_6$ symmetric lattices, in which case the rotation operator does not require any additional gauge transformations. 
    
Ref.~\cite{Biswas2016} also performs a disclination charge calculation for Landau levels in detail, especially studying the contributions from inter-LL states. The analytical derivations of disclination charge are done in the continuum limit, and numerically verified for lattice systems. The calculations are all performed on closed surfaces with corners, so that the charge of a single disclination is deduced by symmetry, analogous to the $Q_{\text{cube}}$ results discussed in this paper. This work is effectively restricted to parts of the Hofstadter butterfly that reduce to Landau levels in the continuum limit. 
    
Ref.~\cite{Li2020disc} derives a formula for disclination charge when Chern number $C=0$, in terms of the charge distribution at high-symmetry points of the unit cell. Their definition of disclination charge is distinct from ours. In the $C_4$ symmetric case, it is equivalent to choosing a region $W$ in the \textit{dual} lattice, and evaluating $\bar{Q}_W$ for this $W$ as we have outlined. If we denote their disclination charge result as $Q_{\text{disc}}^{\text{TZBH}}$, we find that $\frac{\mathscr{S}}{4} - Q_{\text{disc}}^{\text{TZBH}} = \frac{\nu_0}{4}$ for $C_4$ symmetric lattices with zero Chern number. We note that our definition of disclination charge, $\bar{Q}_W$ with $W$ defined on the physical lattice, is directly proportional to $\mathscr{S}$, which also agrees with the coefficient of the linear contribution to the angular momentum associated to vertex-centred rotations. In this sense $\bar{Q}_W$ as defined in the main text is more directly related to the Wen-Zee coefficient and the prediction from the topological field theory. 

In the $C \neq 0$ case, Ref.~\cite{Li2020disc} writes down the Wen-Zee term with $\mathscr{S}=C$ for specific models. It also proposes a formula for disclination charge in the $C \ne 0$ case, in terms of band invariants at high-symmetry points of the Brillouin zone. The values of disclination charge from this definition are different from ours, although we have not studied this relationship in detail when $C \ne 0$. This paper does not make an explicit connection between this formula and the Wen-Zee coefficient $\mathscr{S}$. In our work we have shown that $\mathscr{S}$, which is proportional to $\bar{Q}_W$, depends sensitively on the Chern number through the relation $\mathscr{S} = \frac{C}{2} \mod 1$.

In related work, Ref.~\cite{2108.00008} discussed a spinless fermion system on the square lattice with $C = 0$, and calculated the fractional charge bound to a disclination at half filling. Finally, Refs.~\cite{shiozaki2017invt,You2020hoe} study the angular momentum response using partial rotations in bosonic and fermionic models respectively. The specific models studied numerically have Chern number $C=0$.

\section{Jumps of $\mathscr{S}(\phi)$}\label{sec:Farey}

\begin{figure}[t]
\includegraphics[width=8.0cm]{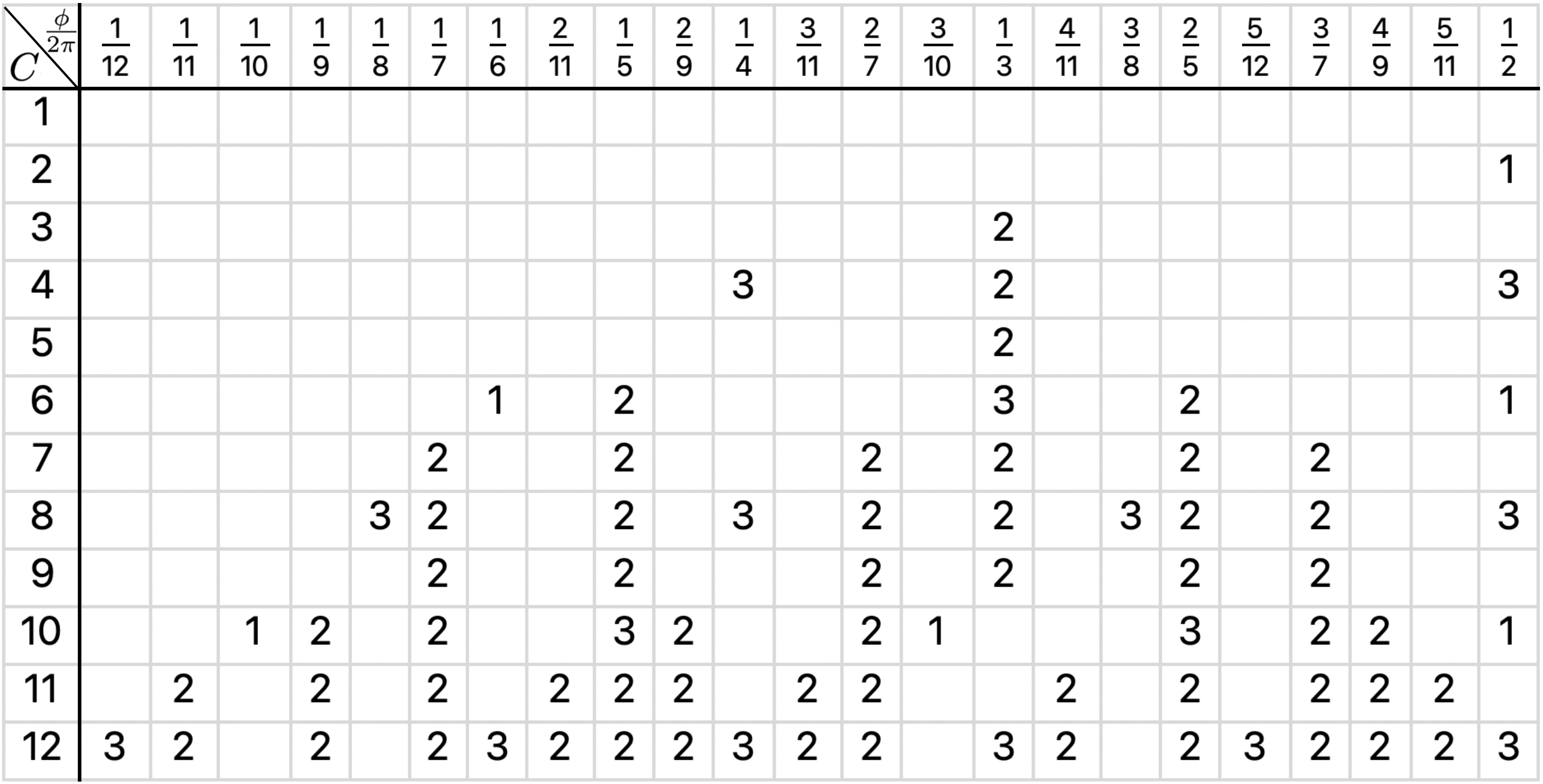}
\caption{\label{fig:farey} Jumps in $\mathscr{S}(\phi)$ for fixed $C$, as $\frac{\phi}{2\pi}$ increases from 0 to $\frac{1}{2}$. }
\end{figure}

In this section we tabulate the data used to empirically obtain Eq.\eqref{eq:s_formula} in the main text. Starting from the zero-flux value $\mathscr{S}(\phi = 0^+) = C^2/2$, Fig.~\ref{fig:farey} shows how $\mathscr{S}(\phi)$ jumps as $\phi$ is increased keeping $C$ fixed. The Hofstadter lobes corresponding to Chern number $C$ are separated by the Farey sequence of order $2|C|$. But from the figure, we see that $\mathscr{S}(\phi)$ only jumps at fractions $\frac{p}{q}$ lying in the Farey sequence of order $|C|$, which is a smaller set. 
We have confirmed this up to $C=12$, see Fig.~\ref{fig:farey}. For example, at $C=4$, the Farey sequence of order 8 is $0,\frac{1}{8},\frac{1}{7},
\frac{1}{6},\frac{1}{5},\frac{1}{4},\frac{2}{7},\frac{1}{3},\frac{3}{8},\frac{2}{5},\frac{3}{7},\frac{1}{2}, \dots$,
with the rest of the sequence obtained by reflecting about $\frac{1}{2}$. However, $\mathscr{S}$ only jumps at $\frac{\phi}{2\pi} = \frac{1}{4},\frac{1}{3},
\frac{1}{2},\frac{2}{3},\frac{3}{4}$, all of which lie in the Farey sequence of order 4.

\section{Definition of disclination Hamiltonian using $\tilde{C}_{4,\lambda}$}\label{sec:Hdisc}
In this section we first define $\tilde{C}_{4,\lambda}$ on the infinite plane; $\lambda$ is fixed up to a constant $\lambda_o$ by the background vector potential $A$, where $o$ is the rotation center. 
We then construct a disclination Hamiltonian $H_{\text{disc}}$ by removing a quadrant and connecting the open edges by defining new hopping terms. This is essentially an adaptation of the prescription to construct symmetry defects for on-site symmetries given in Ref.~\cite{Barkeshli2019}. The values of the new hopping terms are determined by $\tilde{C}_{4,\lambda}$. We show that requiring the disclination to not insert any additional $U(1)$ flux, fixes $\lambda_o=0$. For a fixed $A$ on the infinite plane, this completely fixes $\lambda$.

\subsection{Definition of $\tilde{C}_{4,\lambda}$ and $H$ on an infinite plane}
We define the rotation operator on the infinite plane as
\begin{equation}
    \tilde{C}_{4,\lambda} = \hat{C}_4 e^{i \sum_j \lambda_j c_j^{\dagger} c_j}
\end{equation}
where $\hat{C}_4$ implements a counterclockwise $\pi/2$ spatial rotation about some origin $o$. $\lambda$ depends on our definition of $A$, see below. In what follows we will simplify our notation by using $A_{ij}$ as $A_{ij}=\delta A_{ij}+A_{bgd,ij}$. We also assume that the rotation center at $o\equiv(0,0)$ coincides with the disclination center.

The Hamiltonian $H$ on the infinite plane is given by
\begin{equation}
    H = -t\sum_{<ij>} e^{-i A_{ij}} c_i^{\dagger} c_j + h.c.
\end{equation}
with $i,j \in \Z^2$.
We assume a general vector potential  $A_{ij}$ on the infinite plane that adds constant flux $\frac{\phi}{2\pi}$ per plaquette. $\lambda$ satisfies the equation
\begin{equation}\label{eq:lambdaclean1}
A_{\hat{C}_4i,\hat{C}_4j}=A_{ij}+\lambda_j-\lambda_i
\end{equation}
due to the condition that $\tilde{C}_{4,\lambda}$ commutes with $H$. Therefore, given a gauge choice $A_{ij}$, $\lambda_j$ is completely fixed up to a global constant as follows. Consider a connected path $(j_0\rightarrow j_N)\equiv j_0,j_1,j_2,\dots,j_N$, where $j_0\equiv o$ and $j_N\equiv f$. Let's define $\int_{j_0 \rightarrow j_N} A\equiv \sum_{k=1}^N A_{j_{k-1}j_{k}}$. Then $\lambda_f$ is determined by
\begin{equation}\label{eq:lambdaclean}
    \lambda_f = \lambda_{o}+ \int_{\hat{C}_4 j_0 \rightarrow \hat{C}_4 j_N} A - \int_{j_0 \rightarrow j_N} A.
\end{equation}

\subsection{Definition of $H_{\text{disc}}$ through cut and glue procedure}

\begin{figure}[t]
\centering
 \includegraphics[width=8.2cm]{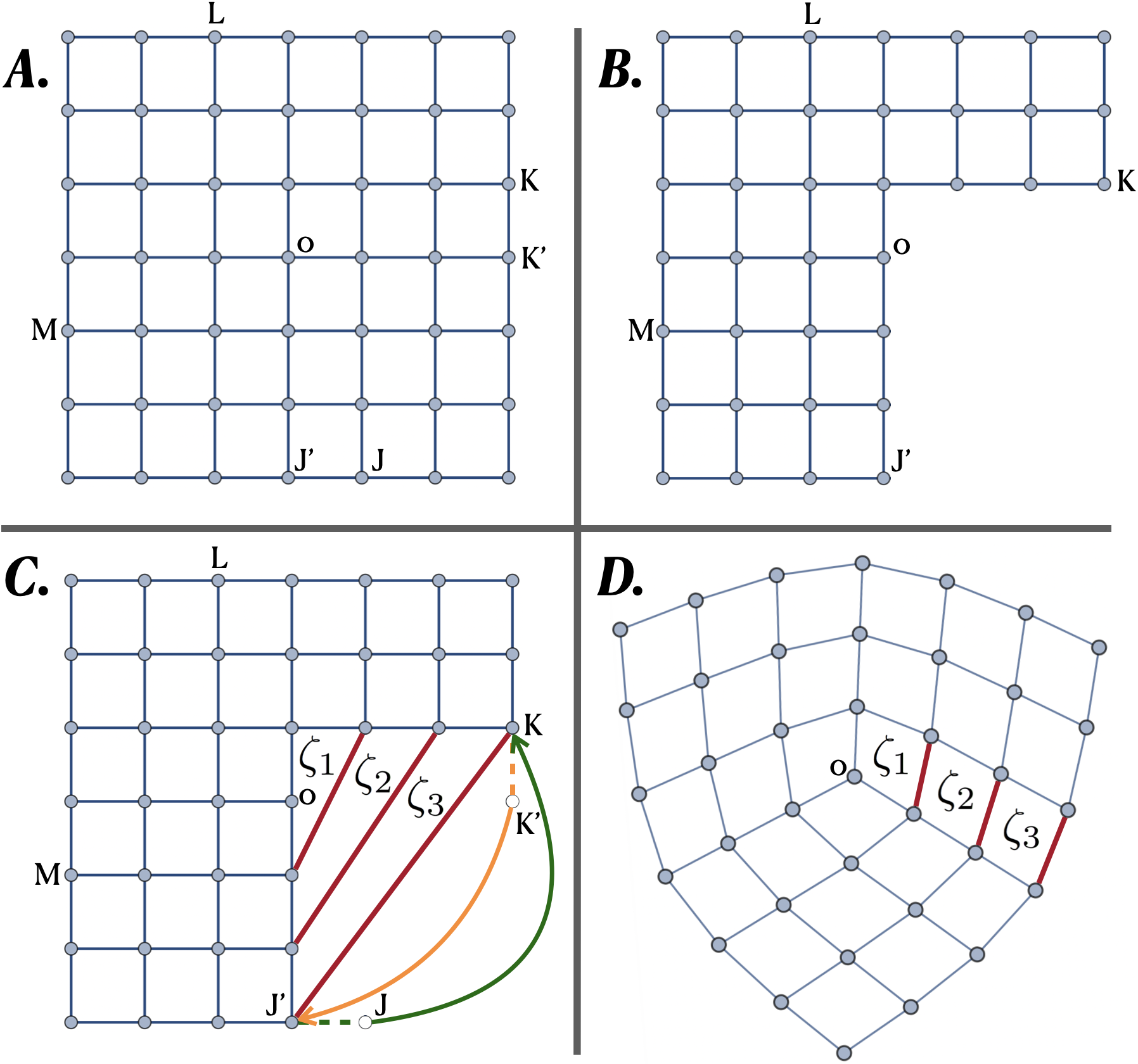}
 \caption{\label{fig:cutNglue} Cut and glue procedure of constructing a disclination. \textbf{A.} Original lattice on an open plane; \textbf{B.} Cutting; \textbf{C.}\xlabel[5C]{fig:cutNglueC} Gluing, with two different ways (green and orange path) of constructing the red bonds. This creates new plaquettes $\zeta_i$; \textbf{D.}\xlabel[5D]{fig:cutNglueD} Reorganizing.}
 \end{figure}

Now we discuss how the definition of $\tilde{C}_{4,\lambda}$ is used in constructing a $\pi/2$ disinclination. 
We first cut out a $\pi/2$ quadrant of the infinite plane, centred at point $o$. In general, the quadrant removed need not have edges parallel to the $x$ and $y$ directions. For ease of visualization, we show an example in Fig.~\ref{fig:cutNglue} where we cut out the bottom right quadrant, by deleting all points $(x,y)$ with $x > 0, y \le 0$. In this case, the Hilbert space of the disclination Hamiltonian $H_{\text{disc}}$ only includes the local Hilbert spaces at sites $(x,y) \in \Gamma = \Z^2 \backslash \{(x,y)| x>0, y \le 0\}$. 

The cutting procedure results in pairs of severed bonds related by $\hat{C}_4$ which should be glued together. Consider one such pair of bonds in $H$ (see Fig.~\nameref{fig:cutNglueC}) involving points $J$ and $K\equiv \hat{C}_4 J$: these are $c_{J^\prime}^{\dagger}c_{J} e^{-iA_{J^\prime J}}$ (green dotted line) and $c_{K}^{\dagger}c_{K^\prime}e^{-iA_{K K^\prime}}$ (orange dotted line). They get severed because points $J$ and $K^\prime$ are removed from $H_{\text{disc}}$. Thus, $J^\prime$ and $K$ should now be joined in $H_{\text{disc}}$ with a hopping coefficient determined by $\tilde{C}_{4,\lambda}$ as follows. 

We start with the term $c_{J^\prime}^{\dagger}c_{J} e^{-iA_{J^\prime J}}$ in $H$, and conjugate \textit{only} the operator $c_J$ with $\tilde{C}_{4,\lambda}$ (green arrow in Fig.~\nameref{fig:cutNglueC}), i.e.
\begin{equation}\label{eq:const1}
    \begin{aligned}
        c_{J^\prime}^{\dagger}c_{J} e^{-iA_{J^\prime J}} &\rightarrow  c_{J^\prime}^{\dagger} \pqty{\tilde{C}_{4,\lambda} c_{J} \tilde{C}_{4,\lambda}^\dagger} e^{-iA_{J^\prime J}}\\&= e^{-i (A_{J^\prime J}+\lambda_J)}c_{J^\prime}^\dagger c_K
    \end{aligned}
\end{equation}
where we have used $\tilde{C}_{4,\lambda} c_J \tilde{C}^\dagger_{4,\lambda}=e^{-i \lambda_J} c_J$. Therefore, the gauge potential for the newly formed bond $J^\prime K$ is given by $\tilde{A}_{J^\prime K}=A_{J^\prime J}+\lambda_J$. 

There is an alternative construction in which we start with the term $c_{K}^{\dagger}c_{K^\prime}e^{-iA_{K K^\prime}}$ in $H$ and conjugate only $c_{K^\prime}$ with $\tilde{C}_{4,\lambda}^\dagger$ (orange arrow in Fig.~\nameref{fig:cutNglueC}). This would give
\begin{equation}
    \begin{aligned}
        c_{K'}^{\dagger}c_{K} e^{-iA_{K' K}} \mapsto& \left(\tilde{C}_{4,\lambda}^\dagger c_{K'}^{\dagger}\tilde{C}_{4,\lambda}\right) c_{K} e^{-iA_{K' K}}\\&= e^{-i (A_{K' K}+\lambda_{J'})}c_{J'}^\dagger c_K
    \end{aligned}
\end{equation}
This procedure gives $\tilde{A}_{J^\prime K}=A_{KK^\prime}+\lambda_{J^\prime}$ which is the same as obtained previously, because of Eq.~\eqref{eq:lambdaclean1}.

    \begin{figure}[t]
\centering
 \includegraphics[width=6cm]{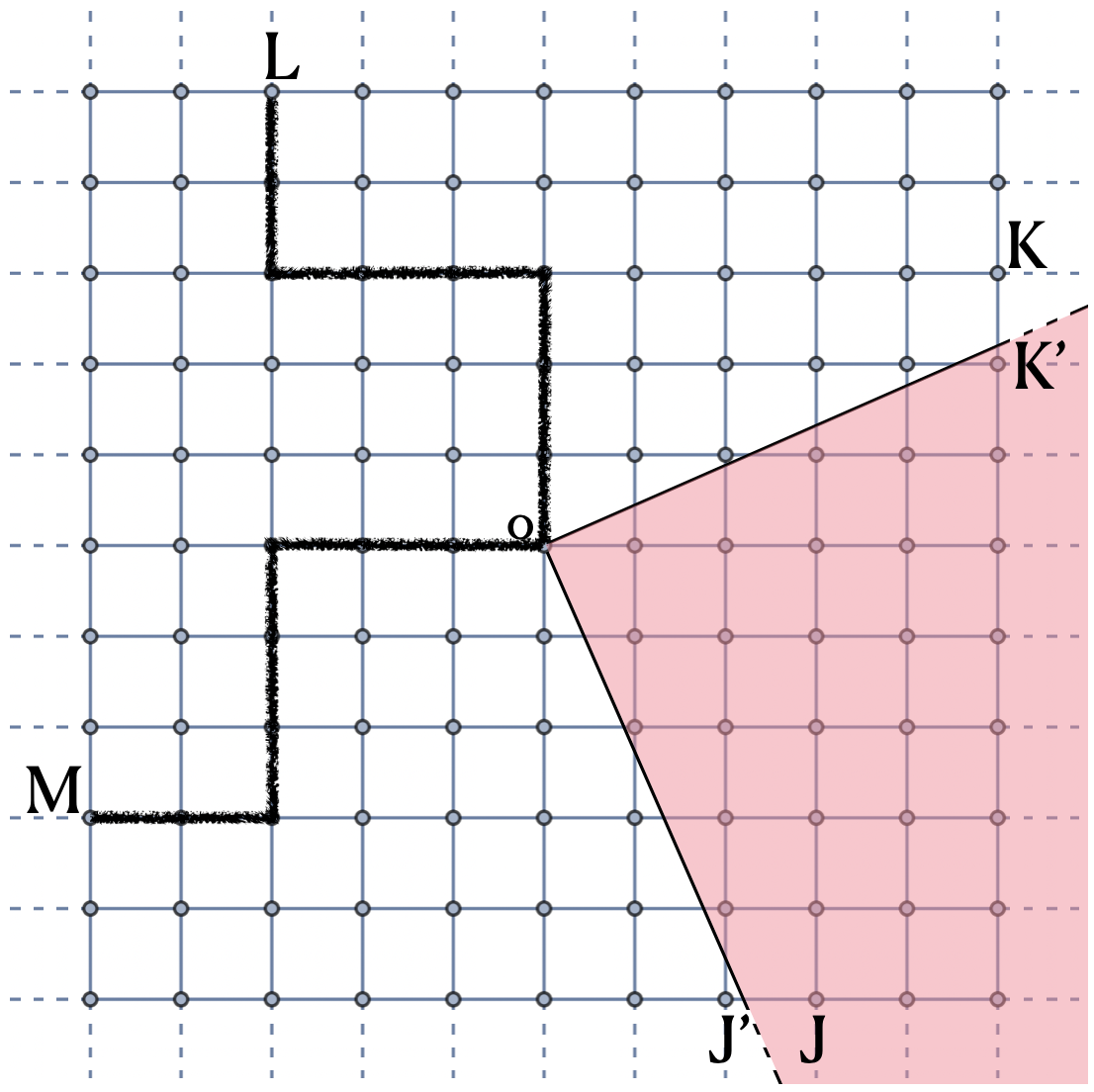}
 \caption{\label{fig:arbregion} A $\hat{C}_4$ symmetric region $D$ with boundary points $J,K,L,M$ that are arranged $\hat{C}_4$-symmetrically about $o$. To construct a disclination, all points in the quadrant shaded red are deleted, and the gluing procedure in Fig.~\ref{fig:cutNglue} is applied.}
 \end{figure}

\subsection{Derivation of $\lambda_o = 0$ on infinite plane}
As we saw above, $\lambda_o$ determines the global $U(1)$ transformation that accompanies spatial rotation. We should keep in mind that conjugating \textit{only} one side of a cut with a global $U(1)$ transformation inserts a $U(1)$ flux through the origin $o$ of the cut. This is the intuition behind why the choice of $\lambda_o$ determines the excess flux $\delta \Phi$ around $o$. Consider a region $D_{\text{disc}}$ that contains a disclination at $o$. It is obtained from a $\hat{C}_4$-symmetric region $D$ on the infinite plane (see Fig.~\ref{fig:arbregion}) by deleting the points in the red quadrant and applying the gluing procedure described above. Then the extra flux at $o$ is
\begin{equation}
\label{eq:delphi}
    \delta \Phi = \Phi_{D_{\text{disc}}}- \frac{3}{4}\Phi_D
\end{equation}
where $\Phi_D$ and $\Phi_{D_{\text{disk}}}$ are the total flux through $D$ and $D_{\text{disc}}$ respectively. We will now show that $\lambda_o=\delta \Phi$. 

Consider 4 points $J,K,L,M$ on the boundary of $D$, such that $J\xrightarrow{\hat{C}_4}K\xrightarrow{\hat{C}_4}L\xrightarrow{\hat{C}_4}M$. We break up the boundary of $D$ into four $\hat{C}_4$ symmetric segments $K\rightarrow L$, $ L\rightarrow M$,  $M\rightarrow N$,  $N\rightarrow J$. Then,
\begin{equation}
\begin{aligned}
\Phi_{D}&=\int_{K\rightarrow L} A + \int_{L\rightarrow M} A + \int_{M\rightarrow J} A + \int_{J\rightarrow K} A\\
\Phi_{D_{\text{disc}}}&=\int_{K\rightarrow L} A + \int_{L\rightarrow M} A + \int_{M\rightarrow J} A + \lambda_J
\end{aligned}
\end{equation}
where we have used Eq.~\eqref{eq:const1} in the second equation. We now use Eq.~\eqref{eq:lambdaclean1} repeatedly to relate each segment to the first segment $K\rightarrow L$:
\begin{equation}
    \begin{aligned}
    \int_{L \rightarrow M} A &=(\int_{K \rightarrow L}A) + \lambda_L - \lambda_K\\
    \int_{M \rightarrow J} A &= (\int_{K \rightarrow L}A) + \lambda_M - \lambda_K\\
    \int_{J \rightarrow K} A &= (\int_{K \rightarrow L}A) + \lambda_J - \lambda_K
    \end{aligned}
\end{equation}
Then, we get $\Phi_D=4(\int_{K\rightarrow L} A)+ \lambda_L +\lambda_M +\lambda_J - 3\lambda_K$ and $\Phi_{D_{\text{disc}}}=3(\int_{K\rightarrow L} A)+ \lambda_L + \lambda_M + \lambda_J -2\lambda_K$. Thus, the extra flux $\delta \Phi$ is (using Eq.~\ref{eq:delphi})
\begin{equation}\label{eq:sumlambda}
   \delta \Phi= (\lambda_J + \lambda_K + \lambda_L +\lambda_M)/4.
\end{equation}
Now we relate $\lambda_J, \lambda_K, \lambda_L, \lambda_M$ to $\lambda_o$ using Eq.~\eqref{eq:lambdaclean}. For this, we use an arbitrary path $o\rightarrow M$ (see Fig.~\ref{fig:arbregion}). Then the other three paths are related via $\hat{C}_4$ as: $o\rightarrow J \equiv\hat{C}_4 (o\rightarrow M) $, $o\rightarrow K \equiv\hat{C}_4 (o\rightarrow J) $ and $o\rightarrow L \equiv\hat{C}_4 (o\rightarrow K) $. So,
\begin{equation}
\begin{aligned}
\lambda_J & = \lambda_o +\int_{o\rightarrow K} A- \int_{o \rightarrow J}A\\
\lambda_K & = \lambda_o +\int_{o\rightarrow L} A - \int_{o \rightarrow K}\\
\lambda_L & = \lambda_o +\int_{o\rightarrow M}A - \int_{o \rightarrow L}A\\
\lambda_M & = \lambda_o +\int_{o\rightarrow N}A - \int_{o \rightarrow M}A\\
\implies &(\lambda_J + \lambda_K +\lambda_L +\lambda_M)/4=\lambda_o.
\end{aligned}
\end{equation}
Therefore 
\begin{equation}
\label{eq:delphilambda}
    \delta \Phi = \lambda_o.
\end{equation}
Demanding $\delta \Phi=0$ fixes $\lambda_o=0$, as desired.

\section{Dependence of $\mathscr{S}$ on definition of rotation operator}
For a fixed $\tilde{C}_{4,\lambda}$, we expect from field theory that we should obtain the same result for $\mathscr{S}$ from either the disclination charge or from the angular momentum. We confirm this in two steps. First we fix $\lambda_o = 0$ on a system with open boundary conditions. We can numerically compute $\mathscr{S}$ using both methods, and the results agree. The second step is to check that $\mathscr{S}$ changes in the same way using either method if we redefine $\lambda$. This is shown below. We redefine $\tilde{C}_4$ by a transformation $\lambda_j \rightarrow \lambda_j + \frac{\pi}{2}\beta$. $\beta$ must be an integer, in order for $\tilde{C}_4$ to have order $4$. We show that it takes $\mathscr{S} \rightarrow \mathscr{S} + C \beta$ in both the disclination charge calculation and clean lattice angular momentum calculation, confirming the duality expected from the field theory.

\subsection{Change in $\mathscr{S}$ from disclination charge calculation}
Applying Eq.~\eqref{eq:delphilambda} to a small region $D$, we see that if $\lambda_o\neq 0$, then there is an extra local flux $\delta \Phi =\lambda_o$ in $D$. (In fact this flux is localized to the plaquette marked $\zeta_1$ in Fig.~\nameref{fig:cutNglueD}).  

Since the system has Chern number $C$, any region $W$ containing this plaquette will have an excess charge $\delta \bar{Q}_W = \frac{\pi \beta}{2}\frac{C}{2\pi}$ that must be attributed to the shift. This means that $\bar{Q}_W = \frac{\mathscr{S}}{4} \rightarrow \frac{\mathscr{S} + C \beta}{4}$. This implies our claim.

\subsection{Change in $\mathscr{S}$ from angular momentum calculation}

Now we show that $\mathscr{S}\rightarrow\mathscr{S}+C \beta$ under $\lambda_j \rightarrow \lambda_j + \frac{\pi}{2}\beta$ in the angular momentum calculation as well. The many-body ground state with Chern number $C$ and flux $\phi + \Delta m/L^2$ per unit cell has $\nu_0L^2 + C \Delta m$ filled single-particle states, with $m = m_0 + \Delta m$ total flux quanta. Redefining the rotation operator by $\beta$ would then shift $l$ to
\begin{equation}
    l(m)\rightarrow l(m)+(\nu_0L^2 + C \Delta m)\beta \mod 4.
\end{equation}
Recall that
\begin{equation}
    -k_0\equiv\frac{C\phi}{2\pi}-\nu_0=\frac{Cm_0}{L^2}-\nu_0.
\end{equation}
Note that $k_0$ is always an integer. Using $m_0=\frac{\phi L^2}{2\pi}$, we find after a short calculation that
\begin{equation}\label{eq:changeofl}
    l\rightarrow l+C m \beta-k_0L^2\beta\mod 4.
\end{equation}
Since on a torus $L$ is an even integer, the last term vanishes. $\mathscr{S}$ is the coefficient of the linear term in $l(m)$, therefore if $\beta$ is a constant, we would have $\mathscr{S} \rightarrow \mathscr{S} + C \beta$, as before.

\subsection{Field theory analysis}
The same transformation appears in the effective response theory. Let $\mathcal{L} = \frac{C}{4\pi} A dA + \frac{\mathscr{S}}{2\pi} A d\omega$. Take $A \rightarrow A + \beta \omega$. The point of this transformation is that it sends $dA \rightarrow dA + \beta d\omega$. Thus the magnetic flux at the disclination is shifted by an amount proportional to the disclination angle. This interpretation agrees with the microscopic analysis of disclination charge in the preceding paragraphs. We can directly check that this transformation shifts $\mathscr{S} \rightarrow \mathscr{S} + C \beta$ but keeps $C$, the $A dA$ coefficient, fixed.

\section{Formula for disclination charge in a general unit cell configuration}\label{Sec:disc_sublattice}

\begin{figure}[t]
\centering
 \includegraphics[width=0.45\textwidth]{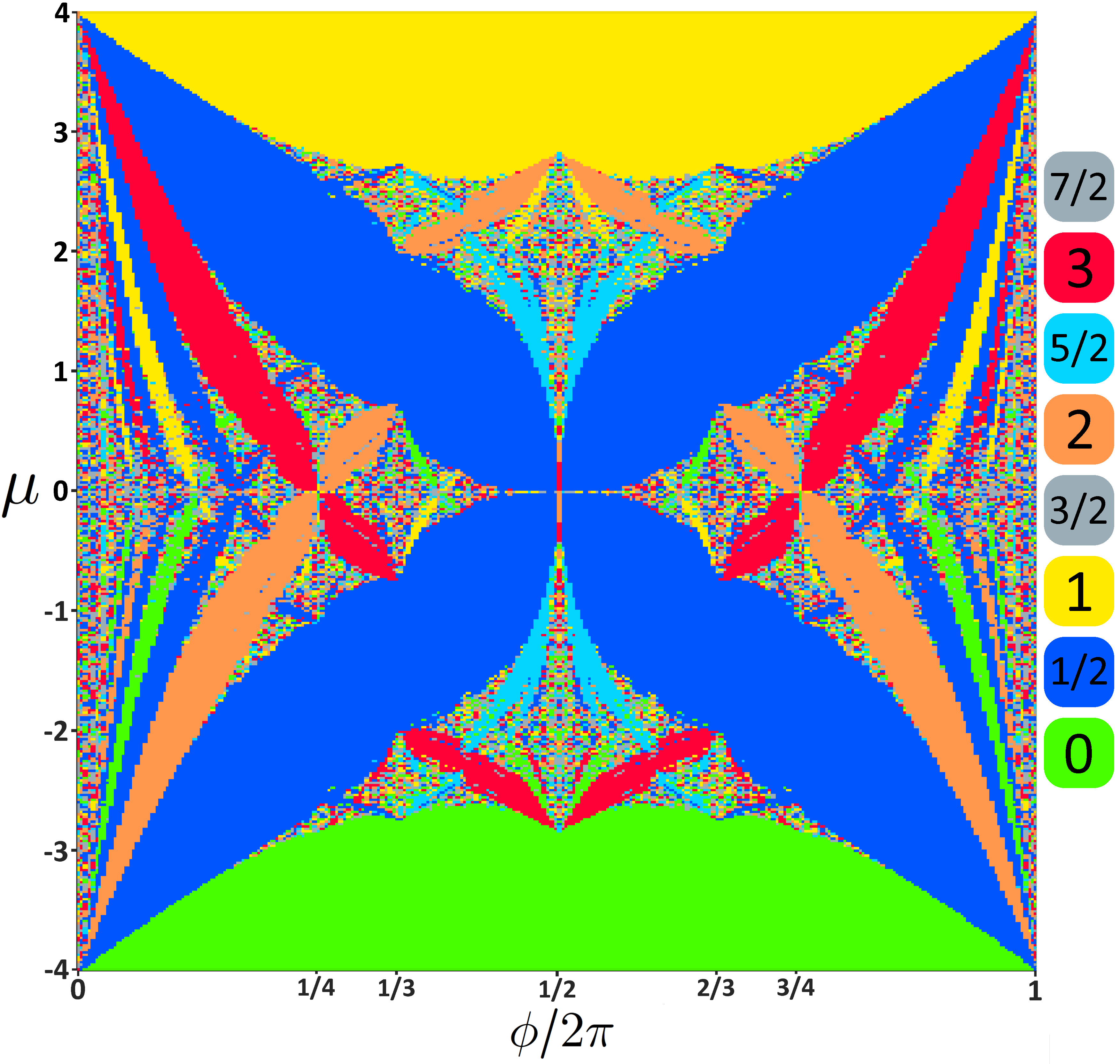}
 \caption{\label{fig:raw_data} Raw data for $\mathscr{S}$ from Eq.~\eqref{eq:Qcube}. The cube has $16$ unit cells per side. The noise in regions with higher Chern number is reduced by taking a larger system size. 
 }
 \end{figure}
 
\begin{figure}[t]
 \includegraphics[width=3.8cm]{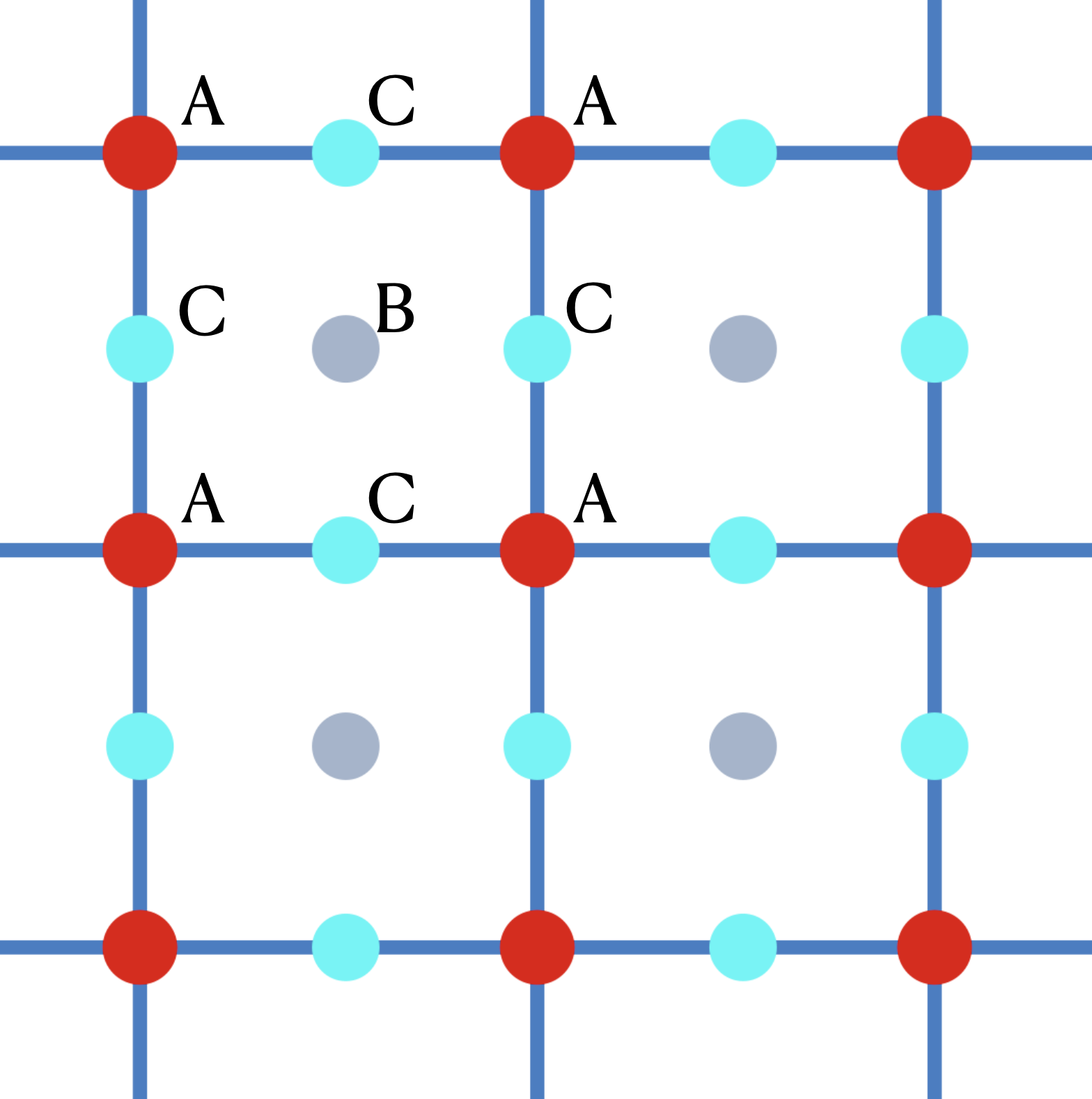}
 \caption{\label{fig:unitcell} General square lattice unit cell with $A$ sites at the corners, $B$ sites inside the unit cell, and $C$ sites on the edges.}
 \end{figure}

Here we assume that the disclination Hamiltonian is already given, and that every unit cell has the same flux. This fixes the rotation operator $\tilde{C}_{4,\lambda}$, as discussed in Appendix \ref{sec:Hdisc}.
 \subsection{Testing $\bar{Q}_W$ for general unit cells}
In the main text, we defined $Q_W = \sum_{i \in W} \text{wt}(i) Q_i$ through Eq.~\eqref{eq:QWfield}. Motivated by the field theory, we chose $\text{wt}(i)$ as in Fig. \nameref{fig:2A}; this led us to the result $\bar{Q}_W \equiv Q_W - \nu_0 n_{u.c.W} = \frac{\mathscr{S}}{4}$ when $W$ encloses a single disclination. Choosing instead $W$ to be the entire surface of a cube gives us Eq.~\eqref{eq:Qcube}, and we plot the raw data of the extracted $\mathscr{S}$ in Fig.~\ref{fig:raw_data}.
In this section we provide some more details on the derivation and its numerical verification, assuming the most general possible unit cell configuration compatible with the square lattice symmetry.

We note that the most general unit cell can be thought of as having $A$ sites on the corners, multiple flavors of $B$ sites within the plaquettes, and multiple flavors of $C$ sites on the edges, see Fig.~\ref{fig:unitcell}. In general there can be several types of $B$ and $C$ sites, but for the purpose of counting charge,
we can treat all sites lying in the interior of the plaquette as a single interior site $B$ lying at the plaquette center.
Similarly, we can treat all sites lying on the edge in the actual microscopic model as a single edge-centered site $C$. We also define the quantities $Q_{\text{refA}},Q_{\text{refB}},Q_{\text{refC}}$, which are the average charges at an $A,B$ and $C$ site respectively far away from any defects. Note that the average charge per unit cell is $Q_{\text{refA}} + Q_{\text{refB}} + 2Q_{\text{refC}} = \nu_0$. This is because each unit cell contains one $A$ site (4 corner sites, each with a weight of 1/4), one $B$ site, and two $C$ sites (4 edge sites, each with a weight of 1/2).

\begin{figure}[t]
\centering
 \includegraphics[width=6.0cm]{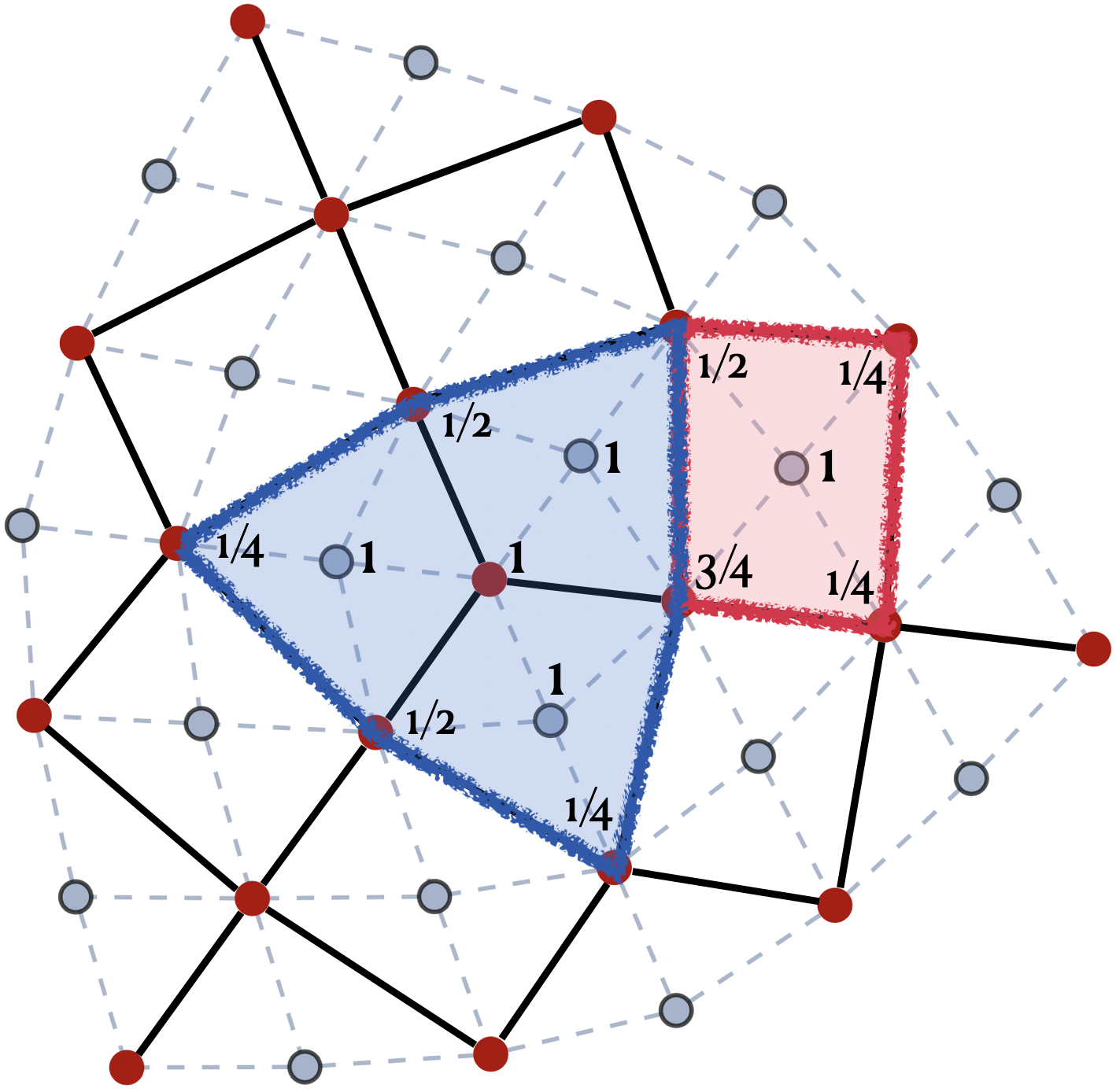}
 \caption{\label{fig:deform} An $AB$ sublattice with red $A$ sites and grey $B$ sites. Black solid lines are unit cell boundaries and blue dashed lines represent hoppings. Adding a new unit cell $Z$ (red region) to $W$ (blue region) will change the weights by 1/4 at each corner of $Z$ and by 1 at the center of $Z$.}
 \end{figure}

Below we show that $\bar{Q}_W = \frac{\mathscr{S}}{4}$ for any region $W$ which encloses a $\pi/2$ disclination, as long as the edge of $W$ is sufficiently far away from the disclination. Let us start by considering a special region $W_0$ such that 8 copies of $W_0$ exactly tile the cube. For this $W_0$, we would have $8 \bar{Q}_{W_0} = \bar{Q}_{\text{cube}} = 2\mathscr{S}$. Thus $\bar{Q}_{W_0} = \frac{\mathscr{S}}{4}$, and this result holds for a general unit cell configuration. Finally we argue that $\bar{Q}_W$ does not change if $W$ is deformed into some region $W'$ which includes an additional unit cell, as long as the deformation occurs far away from the enclosed disclinations. An example of a deformation is given in Fig.~\ref{fig:deform}. Note that $\bar{Q}_{W'} = \bar{Q}_W + \bar{Q}_{Z}$ where $Z$ encloses just one unit cell, and moreover $Z$ is far away from any lattice disclination, so that the $A,B,C$ sites associated to $Z$ have charge $Q_{\text{refA}},Q_{\text{refB}},Q_{\text{refC}}$ respectively. Thus we have $Q_Z = Q_{\text{refA}} + Q_{\text{refB}} +2Q_{\text{refC}} = \nu_0$. As a result $\bar{Q}_Z = Q_Z - \nu_0 = 0$. This implies that $\bar{Q}_{W'} = \bar{Q}_W = \frac{\mathscr{S}}{4}$, i.e. $\bar{Q}_W$ is invariant under deformations of $W$ far away from the disclination. Thus, starting from $W_0$, we can change $W$ while preserving $\bar{Q}_W = \frac{\mathscr{S}}{4}$.  

\begin{figure}[t]
\centering
 \includegraphics[width=9.0cm]{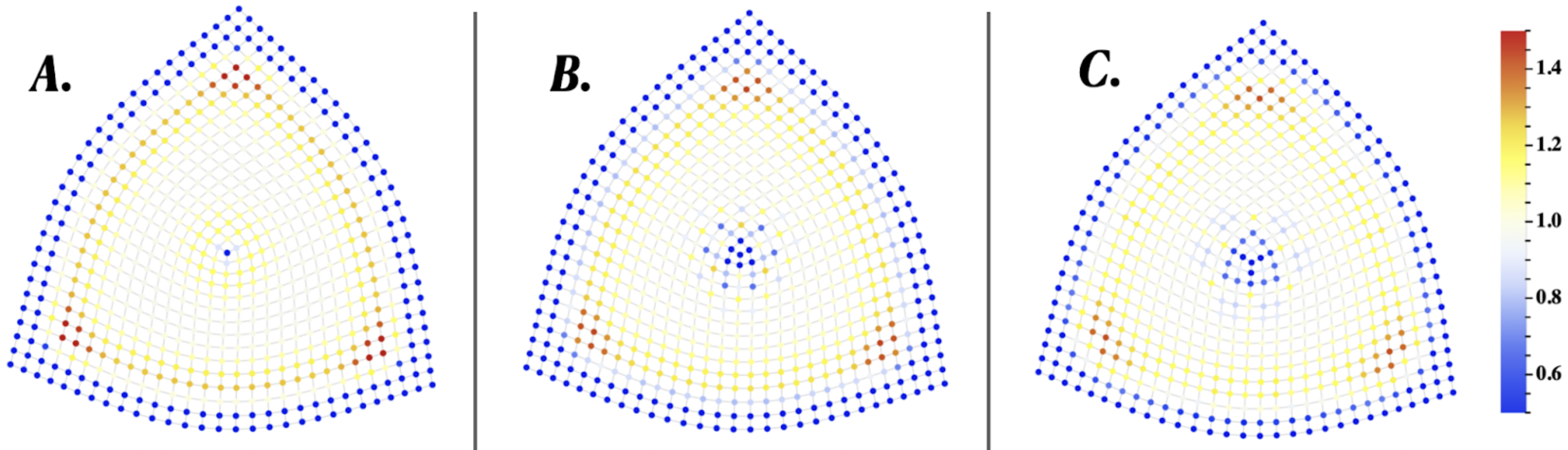}
 \caption{\label{fig:edge} The uncropped charge density distribution of Fig. 2 in the main text. The parameters are \textbf{A.} $\frac{\phi}{2\pi}=\epsilon, C=1$; \textbf{B.} $\frac{\phi}{2\pi}=\frac{1}{2}-\epsilon$, $C=-2$; \textbf{C.} $\frac{\phi}{2\pi}=\frac{1}{3}+\epsilon, C=3$. The colorbar is in units of $\nu_0$. $\epsilon$ is a small fraction which opens up the band gap. The open boundary can host edge states.}
 \end{figure}

If the disclination charge calculation is done on an open disk, we also require that $W$ is defined far away from the boundary. This is because the boundary can host edge states as shown in Fig. \ref{fig:edge}. 

\begin{figure}[t]
\centering
 \includegraphics[width=0.47\textwidth]{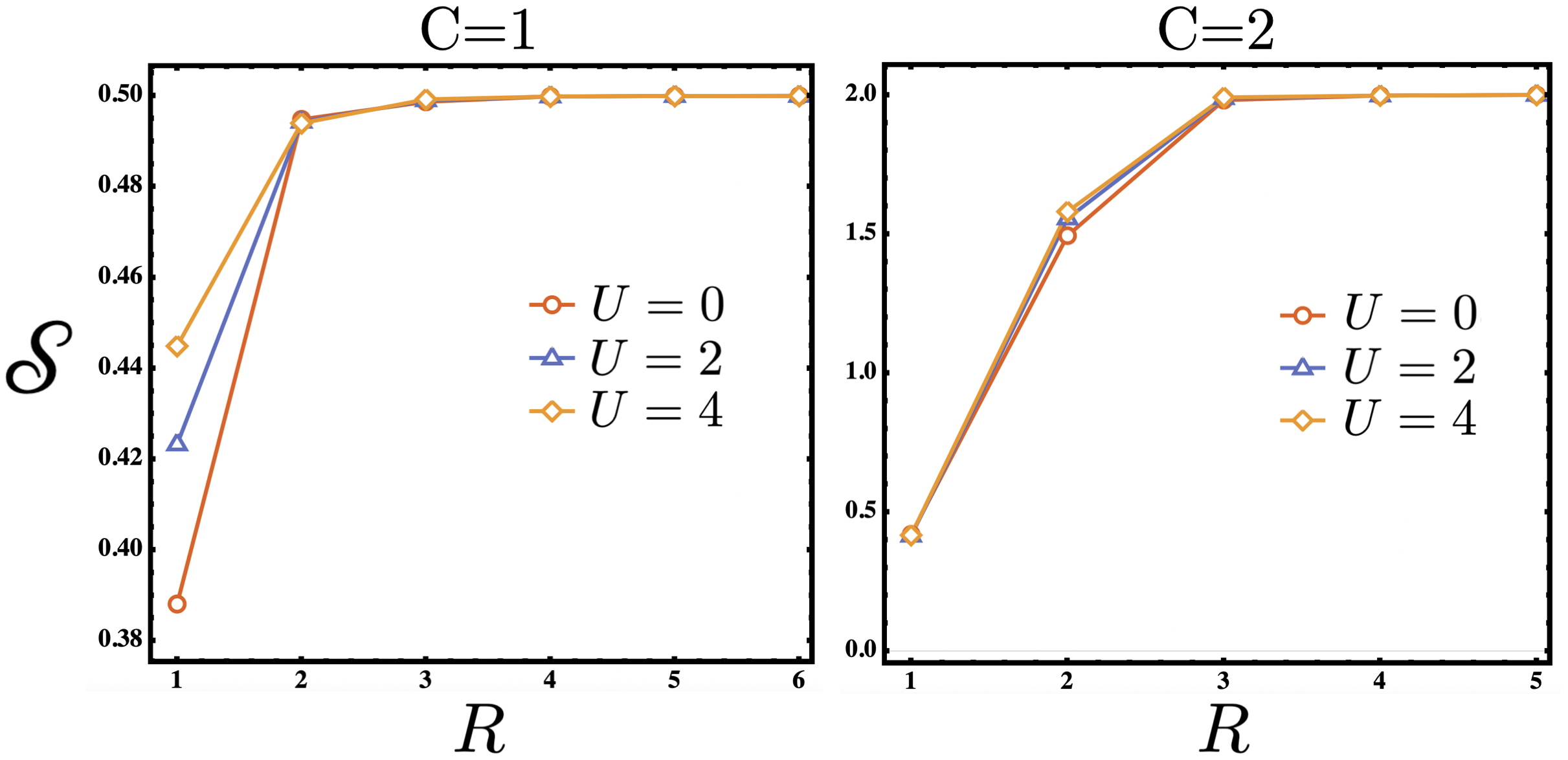}
 \caption{\label{fig:Q_W} Numerical result for $\mathscr{S}$ extracted from $\bar{Q}_W$. Data is taken at $(C,\frac{\phi}{2\pi})=(1,\frac{1}{4})\text{ and }(2,\frac{1}{8})$ in the main Landau levels. We have included different values of an on-site energy $U$ applied only to even ($A$) sites of the square lattice, which gives the square lattice an $AB$ sublattice structure with two sites per unit cell. $W$ encloses a single disclination centered on an $A$ site. $\mathscr{S}$ converges to a quantized value when $R\ge 3$.}
 \end{figure}

In Fig.~\ref{fig:Q_W}, we show the result of our numerical simulations with $AB$ sublattices, i.e. there are no $C$ sites. We consider a cube with 9 unit cells per side. We find that in the main Landau levels, as long as the edge of $W$ is $\approx 3$ lattice units away from the disclination, $\mathscr{S}=4\bar{Q}_W$ is invariant and equals $\frac{C^2}{2}$.

\subsection{Lieb lattice}

\begin{figure}[t]
    \centering
    \includegraphics[width=0.45\textwidth]{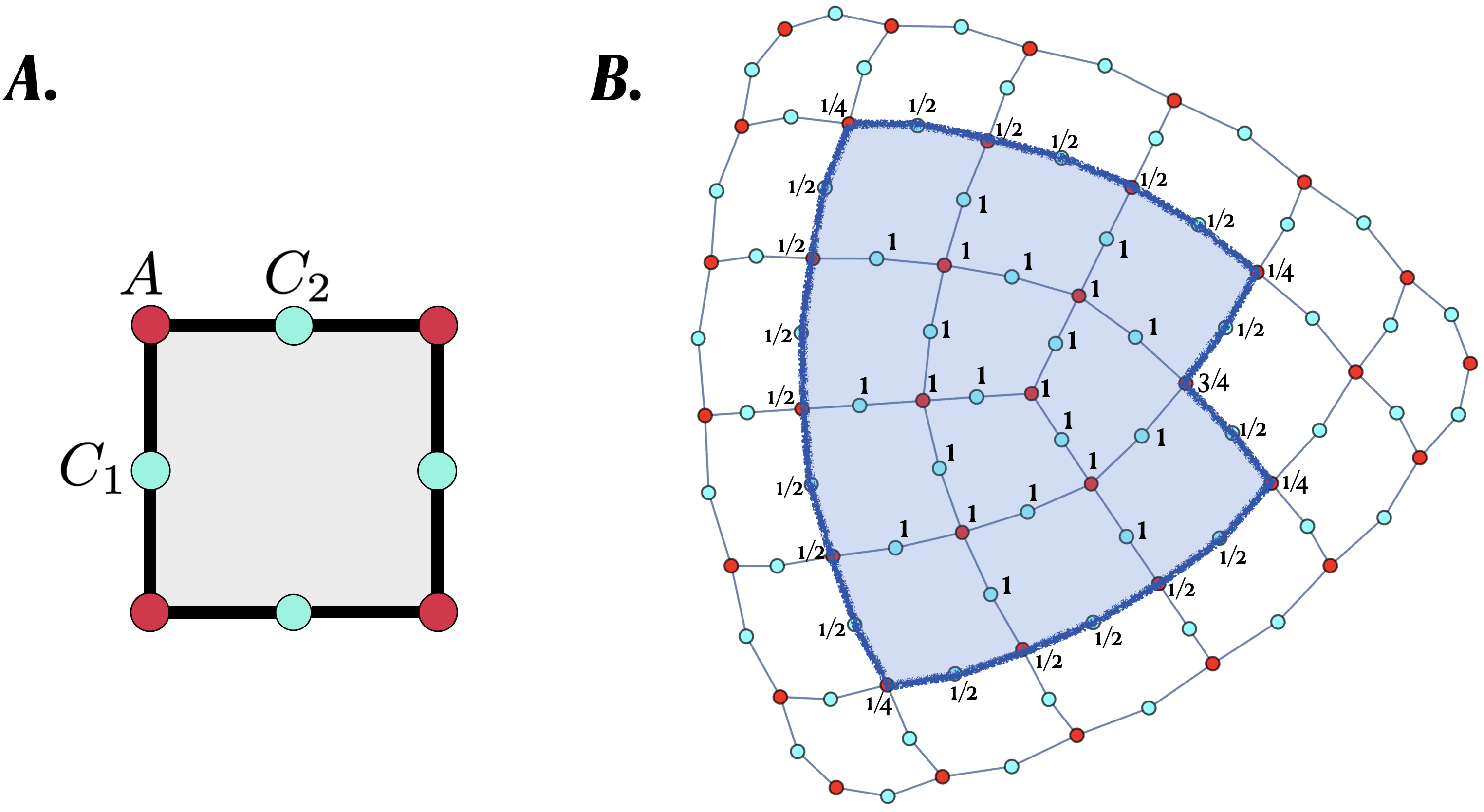}
	
    \caption{\textbf{A.}\xlabel[12A]{fig:liebLA} $C_4$ symmetric unit cell for the Lieb lattice. Solid lines represent hoppings (with unit amplitude) as well as unit cell boundaries. 
    \textbf{B.}\xlabel[12B]{fig:liebLB} $\Omega=\frac{\pi}{2}$ pure disclination for the Lieb lattice. The blue region represents $W$, and the weightings $\text{wt}(i)$ are marked on each relevant site.}\label{fig:liebL}
\end{figure}

In this subsection we use the Hofstadter model on the Lieb lattice as an example to show that our result generalizes to arbitrary $C_4$ symmetric lattices. We pick the $C_4$ symmetric unit cell of the Lieb lattice shown in Fig.~\nameref{fig:liebLA}, where there is one $A$ site and two $C$ sites. A disclination in the Lieb lattice is shown in Fig.~\nameref{fig:liebLB}, and there is a background flux $\phi$ through each unit cell. There is no on site potential on any sites.

\begin{figure}[t]
    \centering
    \includegraphics[width=0.45\textwidth]{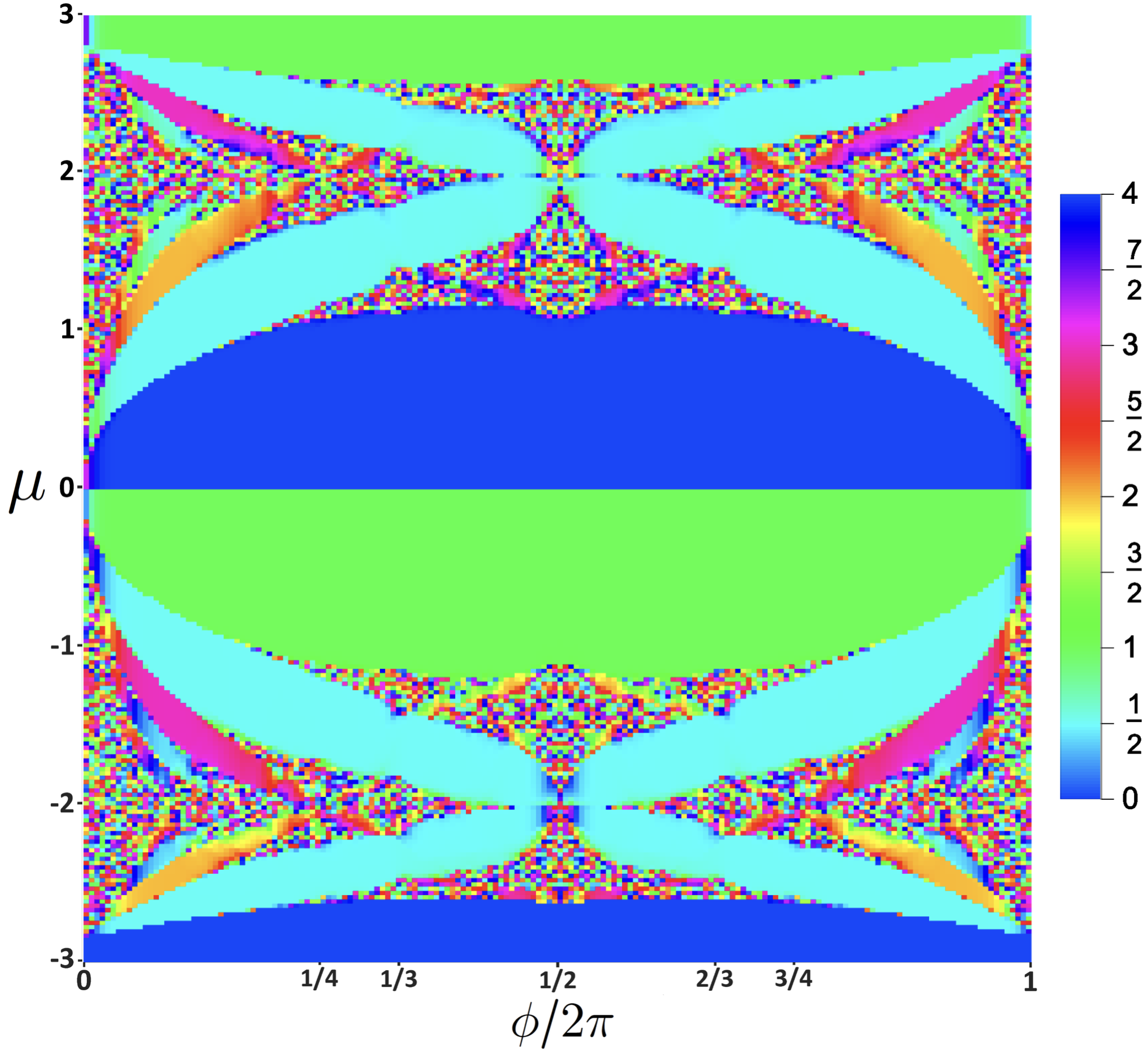}
	
    \caption{Numerical data for the Lieb lattice: $\mathscr{S}$ is extracted using Eq.\eqref{eq:qw}. The calculation is done on an $R=18$ open disk ($R$ is the radius in units of lattice constant). The region $W$ is chosen to be an open disk with $R=9$.  
    }\label{fig:liebS}
\end{figure}

We define $W$ such that the boundary of $W$ is aligned with the boundary of the unit cell. As defined in the main text, the weighting is $\text{wt}(i)=1$ for interior points and $\text{wt}(i)=\frac{1}{4},\frac{2}{4},\frac{3}{4}$ if the interior of $W$ subtends an angle $\frac{\pi}{2},\pi,\frac{3\pi}{2}$ at site $i$. For a disclination on a Lieb lattice, this is explicitly shown in Fig.~\nameref{fig:liebLB}. We can again use the equation 
\begin{equation}\label{eq:qw}
    \bar{Q}_W=Q_W-\nu_0n_{u.c.W}=\frac{\mathscr{S}}{4}
\end{equation}
to extract $\mathscr{S}$. An explicit numerical calculation gives the $\mathscr{S}$ butterfly shown in Fig.~\ref{fig:liebS}.

\subsection{Alternative definition of disclination charge}

It is also natural to use an alternative definition of the disclination charge, different from the one used in the main text, as follows. 

Consider a general unit cell configuration in which the disclination is centered at an $A$ site. 
We define $\check{W}$ to contain a subset of sites on the lattice surrounding the disclination with an extra condition that the number of $A,B,C$ sites in $\check{W}$ excluding the disclination $A$ site be in the same ratio as the number of $A,B,C$ sites in a single unit cell. We do not require $\check{W}$ to enclose an integer number of unit cells. 

\begin{figure}[t]
\centering
 \includegraphics[width=6.5cm]{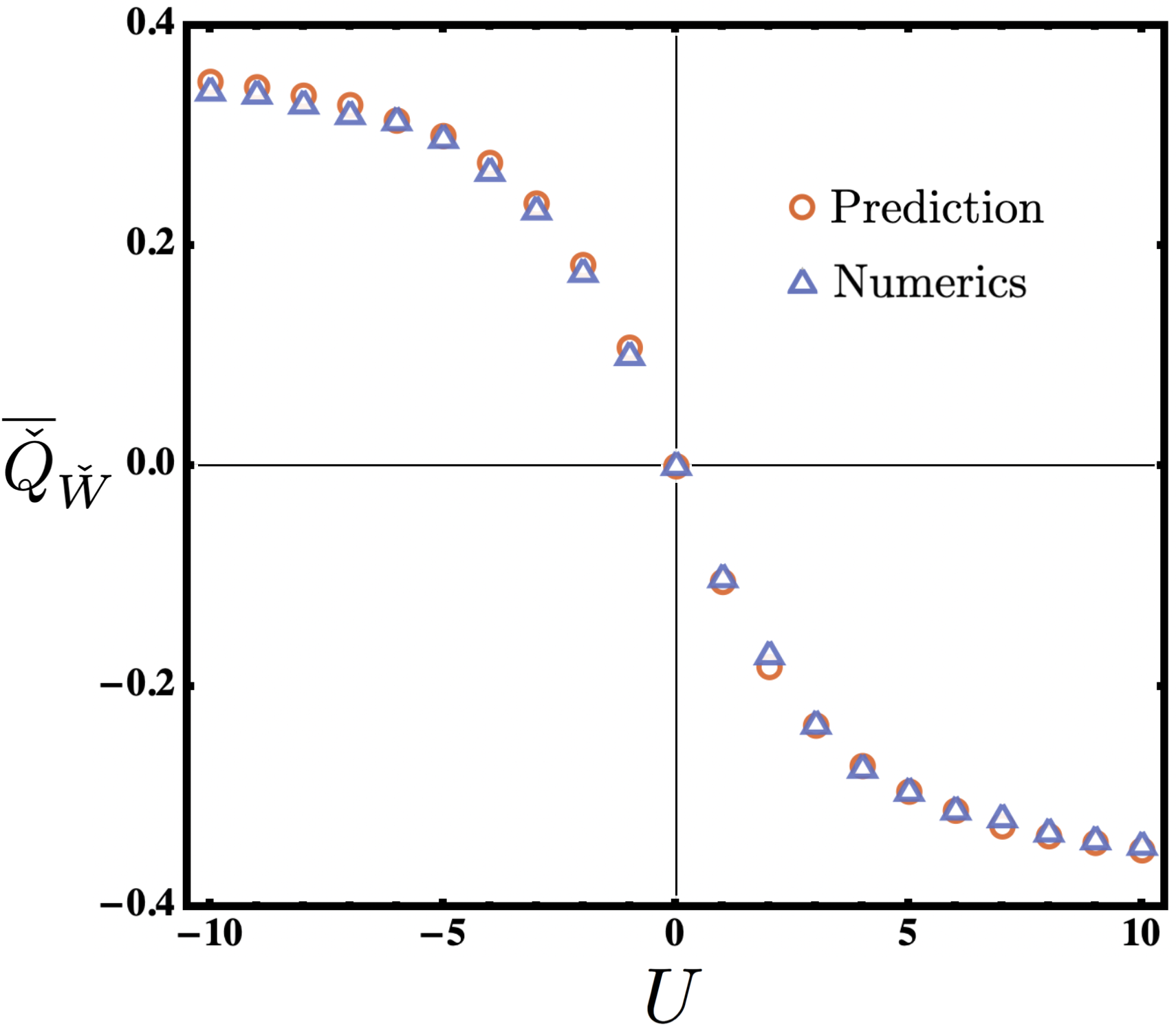}
 \caption{\label{fig:absubQ} Numerical result for $\overline{\check{Q}}_{\check{W}}$ with $\check{W}$ enclosing a $\pi/2$ disclination along with the prediction from Eq.~\eqref{eq:QMbar}. We use an $AB$ sublattice, setting $C=1, \frac{\phi}{2\pi}=\frac{1}{2}-\epsilon$, and on-site energy $-10<U<10$. $\epsilon$ is a small fraction which keeps the band gap open.}
 \end{figure}

We also propose a different definition of charge, namely 
\begin{equation}
    \check{Q}_{\check{W}} \equiv \sum_{i \in \check{W}} Q_i,
\end{equation}
where the weight $\text{wt}(i)=1$ for all sites $i \in \check{W}$.
Since the number of unit cells inside $\check{W}$ is not necessarily an integer, we cannot define a normalized charge $\bar{Q}_W$ as in the main text. Instead, we define the quantity $\overline{\check{Q}}_{\check{W}}$:
\begin{equation}
    \overline{\check{Q}}_{\check{W}}\equiv \check{Q}_{\check{W}}-n_{s,\check{W}}Q_{\text{ref}},
\end{equation}
where $n_{s,W}$ counts the number of sites in $W$ and $Q_{\text{ref}}=\frac{\nu_0}{k_{u.c.}}$ is the average charge per site on a clean lattice. $k_{u.c.}$ is the number of sites inside a unit cell. Using this definition, we can use symmetry and counting arguments to evaluate $\overline{\check{Q}}$. We find that if $\check{W}$ encloses a $\frac{\pi}{2}$ disclination, 
\begin{equation}
    \overline{\check{Q}}_{\check{W}} =\frac{ \mathscr{S}}{4}-\frac{\nu_0}{k_{u.c.}}+\frac{3}{4}Q_{\text{ref}A}, \label{eq:QMbar}
\end{equation}
and this agrees with numerical checks (see Fig.~\ref{fig:absubQ}).

It is more complicated to extract $\mathscr{S}$ using this equation since it also requires us to know $Q_{\text{ref}A}$. Combined with the fact that $\check{Q}_{\check{W}}$ is not additive in the present case, in contrast to what we expect from the field theory, we adopted the definition $Q_W$ instead in the main text.

\section{Angular momentum calculations}\label{sec:AMappendix}

\subsection{Choice of $\lambda$ on the torus}

In Appendix \ref{sec:Hdisc}, we argued that we have to define the magnetic rotation operator with a gauge transformation $\lambda$ that vanishes at the origin, which is the fixed point of the rotation. This choice ensures that the disclination Hamiltonian has the same flux in each unit cell. Now on the torus, there are two fixed points $o_1$ and $o_2$. Thus we have two choices for the definition of the global rotation operator, corresponding to setting $\lambda_{o_1} = 0$ or $\lambda_{o_2} = 0$.

Empirically, we find that if we pick one of the two choices discussed above, $\lambda_{o_1} = 0$ or $\lambda_{o_2} = 0$, then the coefficient of the linear term in $l(m)$ gives an $\mathscr{S}$ that is consistent with the fractional disclination charge result and with Fig. \ref{fig:shift}. 

Note that we can in particular consider redefining $\tilde{C}_4$ with $\lambda_j\rightarrow \lambda_j\pm\frac{\pi}{2}\beta(m)$ with $\beta(m)=-m$. This converts between the two natural choices $\lambda_{o_1} = 0$ and $\lambda_{o_2} = 0$. Using Eq.~\eqref{eq:changeofl}, we see that this also changes the quadratic term in Eq.~\eqref{eq:l} from $\frac{Cm^2}{2}$ to $-\frac{Cm^2}{2}$.

\subsection{Partial rotation}

In the main text, we showed that there is a particular definition of the partial rotation operator $\tilde{C}_4|_D$ which gives the result for $\mathscr{S}$ that is consistent with Fig.~\ref{fig:shift}. Now we discuss some additional subtleties that arise only in the definition of $\tilde{C}_4|_D$. 

\begin{figure}[t]
\centering
 \includegraphics[width=5.2cm]{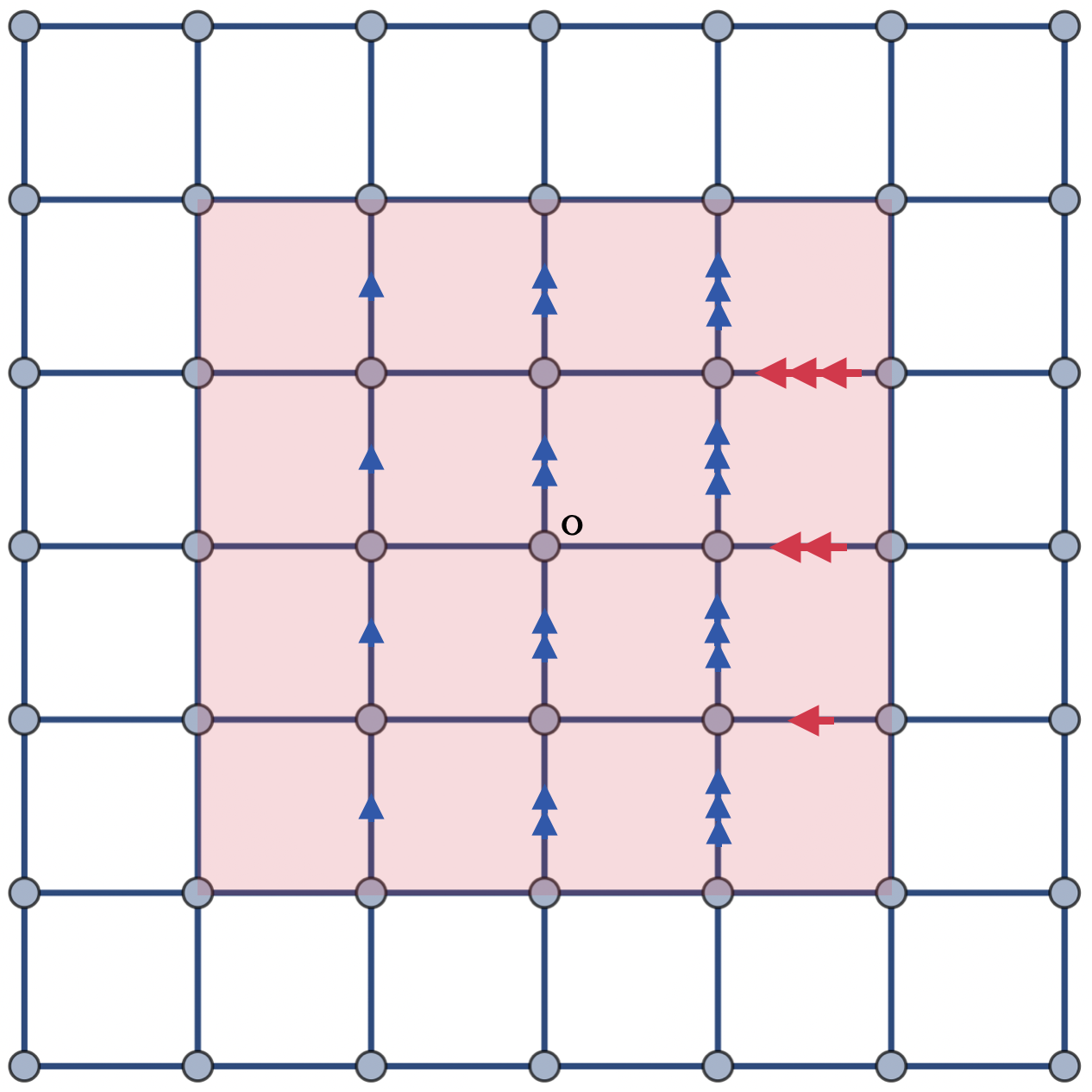}
 \caption{\label{fig:gauge_choice} Insertion of $2\pi\Delta m$ flux distributed uniformly in a local $L'\times L' =4\times 4$ region around the partial rotation center $o$. Each blue arrow represents a hopping phase $A_{ij}=\frac{2\pi\Delta m}{L'^2}$; each red arrow represents a hopping phase $A_{ij}=\frac{2\pi\Delta m}{L'}$.}
 \end{figure}

Suppose we insert $2\pi\Delta m$ flux in a local $L'\times L'$ region $D$. We use a Landau-like gauge, which is shown in Fig.~\ref{fig:gauge_choice}. We can insert the remaining $m_0$ flux quanta globally by inserting $m_0$ flux quanta locally around $o_2$ using this procedure, and taking the limit $L'=L$.

In defining the partial rotation operator, we have freedom in choosing whether $D$ is centered around $o_1$ or $o_2$. In either case, there are two natural choices for $\lambda$, as discussed above. 
Thus in total we have 4 natural choices for $\tilde{C}_{4,\lambda}|_{D}$. We find empirically that only when we enclose $o_2$ with $D$ and define $\lambda_{o_1} = 0$, do we get a shift that is consistent with the global rotation result. In this case, we can calculate $\mathscr{S}$ using the following equation,
\begin{equation}\label{eq:sfroml}
    \mathscr{S} = l_{D}(\Delta m=1, m_0) - l_{D}(0,m_0) - Cm_0 - \frac{C}{2} \mod 4.
\end{equation}

For the other 3 choices, we have not found a meaningful way of extracting $\mathscr{S}$. More specifically, let us assume that $\mathscr{S}$ is given by $\mathscr{S} = l_{D}(\Delta m=1, m_0) - l_{D}(0,m_0) + \sum_j a_j m_0^j \mod 4$. If $a_j$ are arbitrary half-integers independent of $m_0$, we have the identity $\sum_j a_j m_0^j=\sum_j a_j (m_0+8)^j \mod 4$. This requires that $l_{D}(\Delta m=1, m_0) - l_{D}(0,m_0)$ be invariant under $m_0 \rightarrow m_0 + 8$. 
However, empirically we do not observe this when we consider the remaining 3 choices. Thus there appears to be no polynomial fit with half-integer coefficients that can extract $\mathscr{S}$ in these cases.
We do not have a complete theoretical explanation for this, and leave a more detailed study for future work. 

We have tested Eq. \ref{eq:sfroml} on an $L=24$ torus. $D$ is chosen to be a $14\times 14$ square region centered at $o_2$ and the flux is inserted uniformly in a $6\times 6$ region. The resulting values of $\mathscr{S}$ match Fig.~\ref{fig:shift}. 

\subsection{Partial rotation on space with open boundary conditions}

We can repeat the partial rotation calculations on a system with open boundaries. Now there is only one fixed point of rotation $o$. Na\"ively we might wish to take $\lambda_o=0$ for all $m$. However this choice does not work, since we are unable to obtain a uniform $\mathscr{S}$ in a single lobe using a polynomial fit defined as above. 

Empirically, we find that to obtain a value of $\mathscr{S}$ that is constant within a lobe through the above equation, we have to pick the choice $\lambda_o=\frac{\pi}{2}m_0$ before local flux insertion and $\lambda_o=\frac{\pi}{2}m_0+\frac{\pi}{2}\Delta m$  after local flux insertion. 
Using this, $\mathscr{S}$ can be extracted using Eq.~\eqref{eq:sfroml}, and the result matches Fig.~\ref{fig:shift}. This is numerically tested using the same system size dimensions as on the torus. For both open and closed boundary conditions, we have also varied $L,L'$ and the size of $D$. The equation for $\mathscr{S}$ is independent of these parameters as long as they are sufficiently large, and the local flux is completely covered by $D$.

\section{Crystalline gauge theory}
\subsection{Review}\label{sec:CGTreview}

Here we discuss how to write down topological response actions for systems with $U(1)$ charge conservation, discrete magnetic translation and $\Z_M$ rotational symmetries on a spacetime 3-manifold $\mathcal{M}^3$. We do this by defining background `crystalline gauge fields' for the symmetry and then demanding that the action written in terms of these gauge fields obey a topological invariance condition. 

In order to construct actions from discrete gauge fields, it is convenient to work in terms of simplicial cohomology and simplicial calculus (see Appendix A of Ref \cite{Kapustin2014} for a review). The basic idea is to triangulate $\mathcal{M}^3$ and define a discrete background gauge field for the symmetry group $G$ on the links of the triangulation. In the simplicial formulation we demand that the gauge fields be flat; this means that the product of group elements around a 2-simplex (triangle) with 3 edges equals the identity whenever the simplex is contractible. The holonomy of the gauge field through non-contractible cycles can be non-trivial, and the integral of the flux through any closed 2-dimensional submanifold must be quantized. 


To more easily relate our results to the existing literature on the Wen-Zee shift in the context of continuum systems, here we instead consider a continuum formulation. We take the gauge fields to be real-valued differential $1$-forms on $\mathcal{M}^3$. We define 
\begin{align}
\delta A, X, Y, \omega \in \Omega^1(\mathcal{M}^3, \mathbb{R}),
\end{align}
where $\Omega^k(\mathcal{M}^3,\mathbb{R})$ denotes the space of real-valued differential $k$-forms. $\delta A$, $\vec{R} = (X,Y)$, and $\omega$ are the $U(1)$, translation, and rotation components of the full gauge field $B$, now defined as differential $1$-forms. In the continuum formulation, the analog of requiring that the gauge field be flat is that the flux of the gauge fields can have delta function sources of curvature, such that the holonomies always evaluate to trivial group elements in $G$. This mirrors the fact that the flux through a 2-simplex in the simplicial formulation can be an integer multiple of $2\pi$ for a flat gauge field. Next, we impose the same quantization conditions for the holonomies of the gauge field through non-contractible cycles and the total flux through closed 2-manifolds as in the simplicial case.  

We note that the topological action for the background gauge fields is derived by assuming the gauge fields are flat. Nevertheless, we are treating the topological action as a response theory, which determines the universal, long wavelength response of the system to non-flat gauge field configurations. To some extent it is an assumption that we can utilize the topological action for non-flat gauge field configurations and obtain physically correct universal results. Indeed our numerical results involving various types of flux insertion can be thought of as testing the topological response theory beyond the domain where it is originally derived. 

The physical meaning of the gauge fields is the following. As mentioned in the main text, if $A_{bgd}$ is the part of the vector potential which assigns flux $\phi$ to each unit cell, then the total vector potential is given by $A := A_{bgd} + \delta A$, i.e. $\delta A$ is the deviation of the vector potential relative to the uniform background. Physically, $\vec{R}$ measures the number of lattice units traversed along a given path in $\mathcal{M}^3$, while $\omega$ measures the rotation in the local coordinate axes along a path. $\vec{R},\omega$ can thus be identified with the coframe fields and the $SO(2)$ spin connection associated to the spacetime metric on $\mathcal{M}^3$. For a fuller discussion, see Ref.~\cite{manjunath2021cgt}.

Now the triplet $B = (\delta A,\vec{R},\omega)$ is a gauge field for the non-Abelian magnetic space group symmetry. This means that the three components transform according to a specific group law under composition or under gauge transformations. Let ${\bf g} = (e^{2\pi i z},{\bf r},e^{2\pi i \frac{h}{M}})$ be an element of the magnetic space group, where $e^{2\pi i z} \in U(1), {\bf r} \in \Z^2, h \in \Z_M$. The group law is
\begin{widetext}
\begin{equation}
    (e^{2\pi i z_1},{\bf r}_1,e^{2\pi i \frac{h_1}{M}})(e^{2\pi i z_2},{\bf r}_2,e^{2\pi i \frac{h_2}{M}}) = (e^{2\pi i (z_1 + z_2 + \frac{\phi}{2\pi} \frac{{\bf r}_1 \times U(h_1) {\bf r}_2}{2})}, {\bf r}_1 + U(h_1) {\bf r}_2, e^{2\pi i \frac{h_1 + h_2}{M}}),
\end{equation}
\end{widetext}
where $U(h)$ is the $2\times 2$ rotation matrix corresponding to the angle $\frac{2\pi h}{M}$. This group law must be obeyed by $B$. In particular, the integral of $B$ along a cycle $\gamma$ enclosing a region $W$ equals 
\begin{align}
    \oint_{\gamma} B &= \left(\oint_{\gamma} \delta A + \phi \int_W A_{XY}, \oint_{\gamma} U(\smallint_{x_0}^x \omega) \vec{R}, \oint_{\gamma} \omega \right) \\
    &= \int_W \left(d \delta A + \frac{\phi}{2\pi} A_{XY}, d\vec{R} + i \sigma_y \omega \wedge \vec{R}, d\omega\right)
\end{align}

Here the differential 2-form $A_{XY}$ counts the number of unit cells within $W$. When $\omega=0$, $A_{XY} = \frac{1}{4\pi} (X \wedge Y - Y \wedge X)$; when $\omega \ne 0$ it is difficult to write an expression in this continuum formulation, but we can still do so in the simplicial formulation \cite{manjunath2021cgt}. Note that the $U(1)$ component of the above equation measures the total $U(1)$ flux in $W$, because $F := d \delta A + \frac{\phi}{2\pi} A_{XY} = d \delta A + d A_{bgd} = dA$ where $A$ is the full vector potential. For each $x \in \gamma$, $U(\smallint_{x_0}^x \omega)$ represents the net rotation of the coordinate axes between some chosen origin $x_0 \in \gamma$ and $x$, obtained by integrating $\omega$ along $\gamma$. This rotation is then applied to $\vec{R}$ at $x$. $U(\smallint_{x_0}^x \omega) \vec{R}$ is thus a parallel-transport of $\vec{R}$, which accounts for how $\vec{R}$ transforms due to $\omega$. The quantity $\vec{T} := d\vec{R} + i \sigma_y\omega \wedge \vec{R}$ is precisely the torsion 2-form whose integral gives the total Burgers vector of dislocations within $W$. The matrix $i \sigma_y$ is the generator of fourfold rotational symmetries, and appears naturally when $M=4$; if $M$ is changed we must use the corresponding rotation generator in defining $\vec{T}$.

Given any 2-dimensional submanifold $D^2 \subset \mathcal{M}^3$, 
the continuum version of the flatness condition is
\begin{align}\label{eq:fluxquant}
\vec{T}(\vec{x}) &= 2\pi(1-U({\bf h}))\sum_j \vec{l}_j \delta^{(2)} (\vec{x} - \vec{x}_j); \nonumber \\
A(\vec{x}) &= 2\pi \sum_j m_j \delta^{(2)} (\vec{x} - \vec{x}_j); \nonumber \\
\omega(\vec{x}) &= 2\pi \sum_j n_j \delta^{(2)} (\vec{x} - \vec{x}_j)
\end{align}
for $\vec{x}, \vec{x}_j \in D^2$. Importantly, $m_j$, $n_j \in \mathbb{Z}$ and $\vec{l}_j \in \mathbb{Z}^2$. This corresponds to delta function sources at the points $\vec{x}_j \in D^2$ which carry integer multiples of the appropriate flux quanta. $U({\bf h})$ is the matrix generator of $\Z_M$ rotations. The factor of $(1-U({\bf h}))$ in the above equation is imposed to match the simplicial result in which the flux due to a dislocation Burgers vector is treated as trivial if the flux of the form $2\pi (1-U({\bf h})) \vec{n}$, for integer $\vec{n}$ \cite{manjunath2021cgt}. 

To describe non-trivial symmetry defects, we want to allow non-flat gauge field configurations. In the continuum theory, this means we need to allow the total flux associated to the delta function sources above to be suitable fractions of the above flux quanta on open manifolds.  
This is equivalent to saying that
\begin{align}
    \vec{T}(\vec{x}) &= 2\pi\sum_j \vec{l}_j \delta^{(2)} (\vec{x} - \vec{x}_j) \\
\omega(\vec{x}) &= \frac{2\pi}{M} \sum_j n_j \delta^{(2)} (\vec{x} - \vec{x}_j)
\end{align}
There is no constraint on $dA$, since the magnetic flux is $U(1)$-valued.

Even when we allow non-flat gauge field configurations, we still require that the integral through any closed $2$d submanifold $D^2$ is appropriately quantized:
\begin{align}
    \frac{1}{2\pi}\int_{D^2} \vec{T} &=  \sum_j \vec{l}_j = (1-U({\bf h})) \vec{l}_{\text{tot}};
    \nonumber \\
    \frac{1}{2\pi}\int_{D^2} dA  &= \frac{1}{M}\sum_j m_j = m_{\text{tot}}
    \nonumber \\ 
    \frac{1}{2\pi}\int_{D^2} d\omega &= n_{\text{tot}}, 
\end{align}
where $m_{\text{tot}}, n_{\text{tot}} \in \Z$ and $\vec{l}_{\text{tot}} \in \Z^2$.

Gauge transformations correspond to conjugating $B$ by gauge variables $b$, i.e. $B \rightarrow b^{-1} B b$. The gauge transformations are also real-valued. In particular large gauge transformations for $A$ and $\omega$ must be quantized in units of $2\pi$. By construction, the holonomy of $B$ is gauge-invariant. 

We now discuss how to construct topological response theories in terms of $B$. 
\subsubsection{Derivation for bosonic SPTs}
We first show the general calculation for bosonic SPT phases, in which the details are simpler. It is known that bosonic SPTs with symmetry $G$ are classified by the group $\mathcal{H}^3(G,U(1))$, and that any allowed response action is related to a 3-cocycle representative within this group. The response coefficients are quantized by a 3-cocycle condition, along with the flux quantization conditions on the gauge fields introduced above. 

Let $\mathcal{L}$ be a 3-form written in terms of the components of $B$. For bosonic SPT phases, the condition that $\mathcal{L}$ be a topological action is 
that the partition function of the theory be invariant under a cobordism that deforms $\mathcal{M}^3$ to some other 3-manifold $\mathcal{M}'^3$ \cite{Kapustin2014}. This cobordism is given by a 4-manifold $\mathcal{W}^4$ such that $\partial \mathcal{W}^4 = \mathcal{M}^3 \cup \overline{\mathcal{M}'^3}$. We have
\begin{equation}\label{eq:bSPT_top}
    e^{i \int_{\mathcal{M}^3 } \mathcal{L}} = e^{i \int_{\mathcal{M}'^3} \mathcal{L}}.
\end{equation}
when $\mathcal{M}^3, \mathcal{M}'^3$ are cobordant. Using Stokes' theorem, we can write this condition as $e^{i \int_{\mathcal{W}^4} d\mathcal{L}} = 1$. By gluing different manifolds with boundary, we can extend this condition to arbitrary $\mathcal{W}^4$, including closed manifolds. 
We now consider $\mathcal{W}^4 = D_1^2 \times D_2^2$, where $D_1^2, D_2^2$ are both closed. The integrals of $F, \vec{T}, d\omega, A_{XY}$ over $D_1^2, D_2^2$ are all quantized, due to Eq.~\eqref{eq:fluxquant}. By taking products of these 2-forms, we can guess a general expression for $d\mathcal{L}$:
\begin{widetext}
\begin{align}
    d\mathcal{L}_{\text{}} &= \frac{C/2}{2\pi} F \wedge F  + \frac{\mathscr{S}}{2\pi} F \wedge d\omega +\frac{\vec{\mathscr{P}}_c}{2\pi} \cdot F \wedge \vec{T} +\frac{k_{0}}{2\pi} F \wedge A_{XY} \nonumber \\
    &+\frac{\ell_s/2}{2\pi} d\omega \wedge d\omega + \frac{\vec{\mathscr{P}}_s}{2\pi} \cdot d\omega \wedge \vec{T} +\frac{k_{s}}{2\pi} d\omega \wedge A_{XY} +\frac{\Pi_{ij}/2}{2\pi} T_i \wedge T_j +\frac{\vec{k}_p}{2\pi} \cdot \vec{T} \wedge A_{XY} + \frac{\alpha/2}{2\pi} A_{XY} \wedge A_{XY}.
\end{align}
\end{widetext}
The coefficients in the numerator of each term are quantized by virtue of Eqs.~\eqref{eq:bSPT_top} and ~\eqref{eq:fluxquant}. For example, $C/2, \mathscr{S}, \ell_s/2$ must be integers, because the integral of $F$ and $d\omega$ over any closed 2-manifold is always a multiple of $2\pi$. Since the integral of $T$ over any 2-manifold lies in $2\pi (1-U({\bf h})) \Z^2$, we additionally obtain that $\vec{\mathscr{P}}_c, \vec{\mathscr{P}}_s \in (1-U({\bf h}))^{-1} \Z^2$. The remaining coefficients can be quantized by similar arguments. By integrating the above expression,  we obtain the desired response theory

\begin{widetext}
\begin{align}\label{eq:bSPT_EA}
    \mathcal{L}_{\text{}} &= \frac{C/2}{2\pi} A \wedge dA  + \frac{\mathscr{S}}{2\pi} A \wedge d\omega +\frac{\vec{\mathscr{P}}_c}{2\pi} \cdot A \wedge \vec{T} +\frac{k_0}{2\pi} A \wedge A_{XY} \nonumber \\
    &+\frac{\ell_s/2}{2\pi} \omega \wedge d\omega + \frac{\vec{\mathscr{P}}_s}{2\pi} \cdot \omega \wedge \vec{T} +\frac{k_s}{2\pi} \omega \wedge A_{XY} + d^{-1} \left( \frac{\Pi_{ij}}{2\pi} (T_i \wedge T_j) +\frac{\vec{\nu}_p}{2\pi} \cdot \vec{T} \wedge A_{XY} + \frac{\alpha}{2\pi} A_{XY} \wedge A_{XY} \right),
\end{align}
\end{widetext}
The total charge in a region can be obtained by computing $\frac{\delta \mathcal{L}}{ \delta A_0} = \frac{C}{2\pi} d \delta A + \frac{C \phi/2\pi + k_0}{2\pi} A_{XY}$. By identifying the term proportional to $A_{XY}/2\pi$ with the filling per unit cell $\nu_0$, we obtain $\nu_0 = C\frac{\phi}{2\pi} + k_0$. By similarly varying $\mathcal{L}$ with respect to $\omega_0$, we can show that $\nu_s = \mathscr{S}\frac{\phi}{2\pi} + k_s$. Note that this action automatically encodes known relations such as $\nu_0 = C\frac{\phi}{2\pi} \mod 1$, since $k_0$ is an integer. 

Certain choices of the coefficients may be nonzero but topologically trivial. For example, if we choose $\mathscr{S}$ to be a multiple of $M$, the resulting partition function equals 1 for any 3-manifold, and thus only $\mathscr{S} \mod M$ is a topological invariant, classified by the group $\Z_M$. The classification of the other terms is obtained for bosonic symmetry-enriched topological phases in Refs. \cite{manjunath2021cgt,Manjunath2020fqh}.

We can prove that this is the most general action possible for the given magnetic space group symmetry by checking that it describes all elements of the cohomology group $\mathcal{H}^3(G,U(1))$, which classifies (2+1)D bosonic SPT phases. This is verified in Ref. \cite{Manjunath2020fqh} using the simplicial formulation. For example, in bosonic SPTs the three terms in parentheses can be written as coboundaries, so they do not represent nontrivial topological responses. 


\subsubsection{From bosonic to fermionic SPT phases}

For fermionic SPT phases, Eq.~\eqref{eq:bSPT_top} cannot be used; the correct topological invariance condition also needs to account for certain additional topological data related to the fermionic degrees of freedom \cite{barkeshli2021invertible}.  We show the required analysis for a simplified case with only charge conservation and rotational symmetries in Appendix \ref{Sec:Quantization_S}. We have also done the more general calculation for the magnetic space group symmetry, which we do not show here. Below we discuss how the general response theory for fermionic SPT phases compares with the bosonic SPT action in Eq.~\eqref{eq:bSPT_EA}.

The terms in Eq.~\eqref{eq:bSPT_EA} are also present in the fermionic SPT response theory, but the coefficients are quantized differently. For example, the three terms in parentheses are not trivial, since the coefficients can be fractional. However these coefficients are completely fixed if we know the previous seven terms. Moreover, they vanish if we set $X,Y$ to zero, so we ignore them in Eq.~\eqref{eq:chargeresponse}.

We also note that in the bosonic SPT case, $C/2$ and $\mathscr{S}$ were only required to be independent integers. However, in the fermionic case they can both be half-integers. In Appendix \ref{Sec:properties_S} we prove this quantization explicitly.

\subsubsection{Chiral central charge and framing anomaly}
One important additional ingredient in the Hofstadter model is the chiral central charge $c_-$, which is proportional to the thermal Hall conductance (and in our case equals the Chern number $C$). For bosonic and fermionic SPTs, $c_-=0$. However, we need to consider more general fermionic \textit{invertible} phases, in which $c_-$ can be any integer or half-integer; the classification of invertible phases depends on the value of $c_- \mod 8$. A system with nonzero $c_-$ always has a framing anomaly, which must be cancelled by adding an extra term $\mathcal{L}_{\text{anom}}= -\frac{c_-}{96\pi} \text{Tr}\left(\Omega \wedge d\Omega + \frac{2}{3} \Omega \wedge \Omega \wedge \Omega\right)$ to the response theory \cite{Gromov2015,witten1989}. Here $\Omega$ is the Levi-Civita spin connection associated to the metric on $\mathcal{M}^3$. In order to describe the spatial symmetry the rotation gauge field $\omega$ must be pinned to $\Omega$. Therefore we identify the relevant components: $\Omega^a_{0,\mu} = \Omega^0_{b,\mu} = 0$, and $\Omega^1_{2,\mu}=\omega_{\mu}$. The framing anomaly can then be rewritten as $\mathcal{L}_{\text{anom}} = -\frac{c_-}{48\pi} \omega \wedge d\omega$. In Eq.~\eqref{eq:chargeresponse}, we collected the $\omega \wedge d\omega$ terms by defining $\tilde{\ell}_s = \ell_s - \frac{c_-}{12}$. 

Finally, we note that when $c_- \ne 0$, several response coefficients, including $C$ and $\mathscr{S}$, can depend on $c_-$. This gives rise to nontrivial relationships among the coefficients. In particular, in Appendix \ref{Sec:properties_S} we argue that $\mathscr{S} = \frac{C}{2} \mod 1$ for spinless fermions.

\subsection{Deriving the linear term in Eq.~\eqref{eq:l} using crystalline gauge theory}
In the empirical result Eq.~\eqref{eq:l} in the main text, the total angular momentum $l(m)$ of a state with $m$ flux quanta has three terms. Here we show that the first term $\mathscr{S} m$ can be explained entirely within the TFT description, after correctly accounting for the magnetic translation symmetry. 
It is important that $m$ be the \textit{total} number of magnetic flux quanta in the state, including the uniform background flux per unit cell $\phi$ as well as any additional flux. We do not have a complete derivation for the $C m^2/2$ term, although we observe it empirically in our numerics.

In the field theory, $\mathscr{S}$ appears through the term
\begin{equation}
    \mathcal{L}_{\mathscr{S}} := \frac{\mathscr{S}}{2\pi} \omega \wedge dA,
\end{equation}
where $dA=d\delta A + \frac{\phi}{2\pi} A_{XY}$ is the total magnetic flux density.
It is a sum of two terms: (i) the flux of $\delta A$, which measures the excess magnetic flux in addition to the background, and (ii) the uniform background flux, which equals $\phi  \int_D A_{XY}$ within any region $D$. 
Suppose $\int_D A_{XY} = L^2$ and $\int_D d\delta A = \Delta m$, with total flux quanta $m = \frac{\phi L^2}{2\pi} + \Delta m$. The total angular momentum in the ground state then equals 
\begin{equation}
    \int \frac{\delta \mathcal{L}_{\mathscr{S}}}{\delta \omega_0} = \mathscr{S}(\Delta m + \frac{\phi}{2\pi} L^2) = \mathscr{S} m.
\end{equation}
The linear term in Eq.~\eqref{eq:l} can thus be straightforwardly explained by accounting for the magnetic translation symmetry, which couples $\omega$ to the background flux and not just the excess flux.

\section{Derivation of constraints on $\mathscr{S}$}\label{Sec:properties_S}
Here we prove the constraints on $\mathscr{S}$ mentioned in the main text. 
\subsection{Derivation of $\mathscr{S} = \frac{C}{2} \mod 1$ for free fermions}
To argue this result in the free fermion setting, we make the following chain of assertions:
\begin{enumerate}
    \item When $C=0$, $\mathscr{S}$ is an integer. To see this, note that when $C=0$, the system can be adiabatically deformed into a set of localized Wannier orbitals at the high symmetry points of the square lattice. This corresponds to an $ABX$ sublattice configuration with $A$ sites on the vertices of the unit cell, $B$ sites on the face centres and $X$ sites at the edge centres. (We temporarily use the notation $X$ instead of $C$ for edge centred sites to avoid confusion with the Chern number $C$.) We assign the sites a charge $n_A,n_B$ or $n_X$, all of which are crucially integers. We have $\nu_0 = n_A + n_B + 2n_X$. (Note that $\nu_0 = C \frac{\phi}{2\pi} \mod 1$ implies that $\nu_0 \in \Z$ when $C=0$.) 

If we consider a cube with the above unit cell structure, the total number of $A,B,X$ sites equals $n_{u.c.,\text{cube}}+2$, $n_{u.c.,\text{cube}}$, $2n_{u.c.,\text{cube}}$ respectively. Therefore we should have
\begin{align}
    Q_{\text{cube}} &= (n_{u.c.,\text{cube}} + 2)n_A \nonumber \\ &+  n_{u.c.,\text{cube}} n_B + 2n_{u.c.,\text{cube}} n_X.
\end{align}
Comparing with Eq.~\eqref{eq:Qcube}, we see that $\mathscr{S} = n_A$, so it must be an integer for $C=0$. 
    \item When the spinless system is composed of $C$ integer filled Landau levels, $\mathscr{S}=C^2/2$ (see eg. Ref. \cite{Wen1992shift} for a proof). This is indeed an integer or a half-integer when $C$ is even or odd, respectively.
    \item A system with general $C$ can always be thought of as a stack of a system composed of $C$ filled Landau levels with another system that has $C=0$. Since $\mathscr{S}$ is additive under stacking, our main claim follows.
\end{enumerate}
\subsection{Argument for $\mathscr{S} = \frac{C}{2} \mod 1$ for interacting fermions}\label{Sec:Quantization_S}

We now present an argument which is valid for interacting fermion systems, using a more sophisticated theory developed in Ref.~\cite{barkeshli2021invertible} along with the general formalism of crystalline gauge fields reviewed in Appendix \ref{sec:CGTreview}. We explain the result on physical grounds, and then give a more formal argument.

\subsubsection{Physical argument}

We will first derive the result assuming a $\mathbb{Z}_4$ rotational symmetry, and then discuss how to generalize the result to a $\Z_M$ rotational symmetry. Consider a system with $U(1)$ charge conservation and a $\mathbb{Z}_4$ rotational symmetry under which the fermion is spinless (i.e. a $2\pi$ rotation acts on fermions with a plus sign). The symmetry acting on bosons is thus $G_b = U(1) \times \mathbb{Z}_4$; the fermionic symmetry is denoted as $G_f = U(1)^f \times \Z_4$. Our argument relies on gauging the fermion parity and studying the $U(1)$ charge at the $\mathbb{Z}_4$ symmetry defects. Strictly speaking, this argument applies only to defects of internal symmetries. However, we can still make progress using the `crystalline equivalence principle' for fermions (fCEP), discussed in Refs.~\cite{Thorngren2018,debray2021invertible}.

For our purposes, the fCEP states that the classification of Chern insulating phases with the above spatial symmetry is identical to that of Chern insulating phases with an \textit{internal} bosonic symmetry $G_b = U(1) \times \Z_4$, where the fermion now has spin-1/2 under the internal $\Z_4$ (i.e. a complete rotation under $\Z_4$ acts on fermions with a minus sign). That is, $G_f^{\text{int}} = [U(1)^f \times \Z_8^f]/\Z_2$. Therefore we now consider the latter internal symmetry, and use the fCEP to claim that the conclusions we obtain will also hold for the spatial symmetry of interest.

Let us gauge the fermion parity. We obtain a topologically ordered system with symmetry $G_b = U(1) \times \Z_4$: the topological order depends on the Chern number $C = c_-$, where $c_-$ is the chiral central charge \cite{barkeshli2021invertible}. The fact that the fermion has spin-1/2 means that encircling a $2\pi$ flux of the $\Z_4$ symmetry (i.e. a composite of four elementary $\Z_4$ defects) gives a minus sign. This sign can be thought of as the braiding phase of the fermion with a fermion parity flux, which is an anyon in the gauged theory, denoted as $m$. Thus inserting a $2\pi$ defect of $\Z_4$ induces the anyon $m$. If we write the elementary defect as $D$, then we have the fusion rule $D^4 = m$.

Let the $U(1)$ charge at $D$ be denoted as $Q_D = \frac{\mathscr{S}}{4}$, in analogy with the result for the original spatial symmetry. Then, $4 \times \frac{\mathscr{S}}{4} = \mathscr{S}$ must be the $U(1)$ charge of 4 copies of $D$. This in turn equals $Q_m$, up to fusion with a fermion. That is,
\begin{equation}
    \mathscr{S} = Q_m \mod 1.
\end{equation}

The crucial observation is that $Q_m$ depends on the fusion rules of the topological order, which in turn depend on $c_-=C$. When $c_-$ is \textit{odd}, $m \times m = \psi$, therefore $Q_m = \frac{1}{2} \mod 1$. When $c_-$ is \textit{even}, $m \times m = I$ (a trivial bosonic particle, which has even charge), therefore $Q_m = 0 \mod 1$. This implies that 
\begin{equation}
    Q_m = \frac{c_-}{2} = \frac{C}{2} \mod 1,
\end{equation}
from which the claim follows. 

The above argument can be generalized to any even $M$ without any difficulties. However, if $M$ is odd, for example $M=3$, we have the following problem. If $D_1$ is an elementary $\Z_M$ disclination satisfying $D_1^M = 1$, then there is another disclination $D_2 = D_1 \times m$, which is a bound state of $D_1$ with a fermion parity defect ($\pi$ flux). It satisfies $D_2^M = D_1^M \times m^M = m$. We would like to define $\mathscr{S} \mod 1$ as the charge of either $D_1^M = 1$ or $D_2^M = m$. But when $M$ is odd, the theory does not tell us which definition to use in general.

Even though the $M=4$ derivation cannot be used, there are still strong arguments that $\mathscr{S} = \frac{C}{2} \mod 1$ is the correct relation even when $M$ is odd. The argument is based on the result that the shift has a $\Z_M$ classification for each $M$, which can be derived from the general theory of Ref.~\cite{barkeshli2021invertible}

For example, take $M=3$. If the $\Z_3$ rotational symmetry in the microscopic model can be viewed as a subgroup of a $\Z_6$ symmetry, then we must have $\mathscr{S} = \frac{C}{2} \mod 1$ in order to be consistent with the $\Z_6$ result. We can always find a system with this property, for any $C$. Now suppose there is a system $A$ for which $\mathscr{S} = \frac{C}{2} + \frac{1}{2}$. We can stack $A$ with a system $B$ that has Chern number $-C$ and $\mathscr{S} = \frac{C}{2} \mod 1$ (say $B$ has a $\Z_6$ rotational symmetry). The stacked system $AB$ will have Chern number zero but $\mathscr{S} = \frac{1}{2} \mod 1$. But this would imply that there are $6$ distinct values of $\mathscr{S}$ when $C=0$, generated by $AB$. But this is inconsistent with the result that $\mathscr{S}$ has a $\Z_3$ classification. Similar arguments apply to all odd $M$.

Below we formalize the physical argument in the language of topological field theory, assuming $M$ is even. 
\subsubsection{Symmetry definition}
Consider a fermionic system in which the symmetry that acts on the bosonic operators is given by $G_b = U(1)_b \times \Z_M$ for some even $M$. (The subscript $b$ emphasizes that this is a bosonic symmetry.) We choose $M$ even so that the spin of the fermion under rotations is well-defined. For the moment $G_b$ is taken to be an internal symmetry, however we will eventually account for the spatial nature of the $\Z_M$ rotations using the `crystalline equivalence principle' \cite{Thorngren2018}. We do not assume any translation symmetry for simplicity, because it is not required to define $C,\mathscr{S}$. 


The symmetry group $G_f$ which acts on fermionic operators is given by a group extension of $G_b$ by the fermion parity $\Z_2^f$, corresponding to a representative cocycle $\omega_2$ in the group $\HH^2(G_b,\Z_2)$. For $G_b = U(1) \times \Z_M$, we have 
\begin{align}
\HH^2(G_b,\Z_2) = \HH^2(U(1),\Z_2)\times \HH^2(\Z_M,\Z_2) = \Z_2 \times \Z_2. 
\end{align}
Let $\alpha_c \in \HH^2(U(1),\Z), \alpha_s \in \HH^2(\Z_M,\Z)$ be representative cocycles generating their respective groups. Denote each element of $G_b$ by ${\bf g}_i = (e^{2\pi i z_i}, {\bf h}_i)$. Then, the physical system has
\begin{equation}\label{eq:om2def}
    \omega_2({\bf g}_1, {\bf g}_2) := \alpha_c(e^{2\pi i z_1},e^{2\pi i z_2}) + k \alpha_s({\bf h}_1, {\bf h}_2) \mod 2.
\end{equation}
The first term on the rhs indicates that a $2\pi$ rotation in $U(1)_b$ acts trivially on any bosonic operator, but transforms any fermionic operator by a minus sign. The second term indicates that a $2\pi$ rotation in $\Z_M$ acts trivially on any bosonic operator, but transforms any fermionic operator by a sign $(-1)^{k}$. The case $k=0$ corresponds to fermions that are `spinless' under the internal $\Z_M$ symmetry; in this case, $G_f = U(1)^f \times \Z_M$. The case $k=1$ corresponds to `spin-1/2' fermions under the internal $\Z_M$ symmetry; in this case, $G_f = (U(1)^f\times \Z_{2M}^f)/\Z_2$.

\subsubsection{Overview of argument}

The main result of Ref.~\cite{barkeshli2021invertible} is that each invertible phase can be described by a set of data $(c_-, n_1, n_2, \nu_3)$ satisfying various consistency conditions and equivalences. $c_-$ is the chiral central charge, defined mod 8. 

$n_1 \in \HH^1(G_b,\Z_2)$ determines whether a ${\bf g}$-defect localizes a Majorana zero mode. One can show that $n_1$ non-trivial is incompatible with $U(1)^f$ symmetry, and therefore we can set $n_1 = 0$ in the present discussion. 

Let $C^n(G_b,M)$ be the space of $n$-variable functions ($n$-cochains) $f_n: G_b \rightarrow M$. The parameter $n_2 \in C^2(G_b,\Z_2)$ fixes the fusion rules of the symmetry defects in the theory, while $\nu_3 \in C^3(G_b,U(1))$ is analogous to a local counterterm that modifies the defect $F$-symbols of the invertible phase, ensuring that they satisfy the pentagon equation for associativity. It is constrained by $\omega_2, c_-, n_2$.

We will be particularly interested in $\nu_3$, because in the case of bosonic SPT phases, it is known \cite{Barkeshli2019} that the response action is obtained by pulling back $\nu_3$ using the background symmetry gauge field $B$. If we think of $B$ as a map from the space-time manifold $\mathcal{M}^3$ to the classifying space $BG$, and let $B^*$ denote its pullback, then the topological action is determined by the Lagrangian $\mathcal{L} = 2\pi B^* \nu_3$. More concretely, this means that if we triangulate space-time and label the 1-simplices of the triangulation with group elements, the action integrated on a single 3-simplex $\Delta^3$ evaluates to $e^{2\pi i \nu_3({\bf g}, {\bf h}, {\bf k})}$, where ${\bf g}, {\bf h}, {\bf k}$ determine the labelings on the 1-simplices of $\Delta^3$.

For bosonic SPTs we have the condition $d\nu_3 = 0 \mod 1$, which is equivalent to Eq.~\eqref{eq:bSPT_top} and which is the topological invariance condition. Thus knowing $\nu_3$ is equivalent to knowing $\mathcal{L}$. 

For invertible fermionic topological phases, we continue to identify the topological response action in terms of the Lagrangian $\mathcal{L} = 2\pi B^* \nu_3$. This is familiar in the case of the usual Hall response, as we will see below. 

We will make use of the following facts. If we consider our space-time manifold $\mathcal{M}^3$ to be the boundary of a $4$-manifold $\mathcal{W}^4$, i.e. $\partial \mathcal{W}^4 = \mathcal{M}^3$, and we extend the background gauge field $B$ over $\mathcal{W}^4$, then we can write
\begin{align}\label{eq:pb_nu3}
    d \mathcal{L} = 2\pi B^* d \nu_3 . 
\end{align}
In the general theory of invertible fermionic topological phases \cite{barkeshli2021invertible}, $\nu_3$ satisfies (when $n_1 = 0$),
\begin{align}
\label{O4anomaly}
d \nu_3 = \mathcal{O}_4 = \frac{1}{2} n_2 \cup (n_2 + \omega_2) + \frac{c_-}{8} \omega_2 \cup \omega_2 \mod 1. 
\end{align}

\subsubsection{Background gauge fields}

We define the bosonic $U(1)_b$ and $\Z_M$ gauge fields $\delta A,\omega$ as in Appendix \ref{sec:CGTreview}. The response theory in the fermionic case can be written in terms of $G_b$ gauge fields if we first gauge the fermion parity symmetry and then consider the resulting anyon theory, which now has $G_b$ symmetry \cite{barkeshli2021invertible}. In addition to $A,\omega$, this gauged theory has `1-form' symmetries which are generated by the fermions and the fermion parity fluxes. This simply means that the closed fermion loops and fermion parity flux loops can be smoothly deformed without affecting the topological path integral. 

We define a background 2-form $\Z_2$ gauge field $B_2^{(2)}$ for the 1-form symmetry generated by the fermion parity fluxes. This means that to each 2-simplex of the triangulated space-time manifold we assign an element of $\Z_2$; if $B_2^{(2)}$ is non-trivial on a 2-simplex, this means that a fermion parity flux is piercing the 2-simplex. This in turn means that a fermion traversing the boundary of such a 2-simplex will acquire a minus sign. Thus, $B_2^{(2)}$ encodes $\omega_2$: if we consider a triangular spacetime region $D \in \mathcal{M}^3$ with gauge fields corresponding to the group elements ${\bf g}_1, {\bf g}_2, ({\bf g}_1 {\bf g}_2)^{-1}$ on the sides, the integral of $B_2^{(2)}$ over $D$ equals $\pi \omega_2({\bf g}_1, {\bf g}_2)$, i.e. 0 or $\pi$. A fermion traversing the boundary of $D$ acquires a sign $(-1)^{\omega_2({\bf g}_1, {\bf g}_2)}$, as expected.  

The definition of $\omega_2$ implies that
\begin{equation}\label{eq:pb_om2}
    B_2^{(2)} = \frac{1}{2} (dA + k d\omega) = \pi B^* \omega_2.
\end{equation}
We write $B_2^{(2)} = \pi B^* \omega_2$ to emphasize that $B_2^{(2)}$ encodes $\omega_2$  defined in Eq.~\eqref{eq:om2def} in terms of $B$.

There is also a 1-form symmetry generated by the fermion, which encodes some additional data $n_2 \in C^2(G_b,\Z_2)$. Proceeding as above, we define a 2-form gauge field 
\begin{align}\label{eq:pb_n2}
B_2^{(1)} = \pi B^* n_2 
\end{align}
for this 1-form symmetry, which encodes $n_2$. Physically, the triangular region $D$ introduced above hosts an extra fermionic degree of freedom when $n_2({\bf g}_1,{\bf g}_2) = 1$ \cite{Wang2020fSPT}.

\subsubsection{Quantization of $\mathscr{S}$}

The most general topological field theory for invertible fermion phase with symmetry $G_f$ (with either choice of $k$) is defined as 
\begin{equation}
    \mathcal{L} = \frac{C}{4\pi} A \wedge dA + \frac{\mathscr{S}}{2\pi} A \wedge d\omega + \frac{\ell}{4\pi} \omega \wedge d\omega
\end{equation}
for some coefficients $C,\mathscr{S},\ell$ to be determined. As discussed above, the main conceptual step is to relate $\mathcal{L}$ to $\nu_3$, by defining $\mathcal{L} := 2\pi \times B^* \nu_3$, where $\nu_3 \in C^3(G_b,\mathbb{R}/\mathbb{Z})$. Assuming this, the quantization of $C,\mathscr{S}$ and $\ell$ is fixed by $c_-$, as we now discuss. For an invertible fermion phase with $U(1)^f$ symmetry, we must have (i) $c_-$ be an integer, and (ii) no Majorana zero modes in the system, i.e. $n_1 = 0$. These two conditions force $dn_2 = 0 \mod 2$ (see Ref.~\cite{barkeshli2021invertible} for a proof). 

The most general choice of $n_2$ is then 
\begin{align}
   n_2 &:= k_c \alpha_c + k_s \alpha_s  \nonumber \\
    \implies B_2^{(1)} &:= \frac{1}{2} (k_c dA + k_s d\omega)
\end{align}
for some $k_c,k_s \in \{0,1\}$. Since there is an equivalence $n_2 \sim n_2 + \omega_2$ in the theory \cite{barkeshli2021invertible}, we can set $k_c = 0$ without loss of generality. 

To constrain $\mathcal{L}$ we now use Eqs.~\eqref{eq:pb_nu3},~\eqref{O4anomaly},~\eqref{eq:pb_om2},~\eqref{eq:pb_n2}. We then have
\begin{align}
     \frac{d\mathcal{L}}{2\pi} &= \frac{1}{\pi^2}\left( \frac{1}{2} B_2^{(1)} \wedge (B_2^{(1)} + B_2^{(2)}) + \frac{c_-}{8} B_2^{(2)} \wedge B_2^{(2)}\right) \nonumber \\ &\mod 1.
\end{align}
(For arbitrary $G_f$, there is an additional term \cite{barkeshli2021invertible}, however for our choice of $G_f$, this term is trivial.) Roughly speaking, this formula indicates that the full theory is topologically invariant under deformations of $\mathcal{M}^3$ if the transformation of the counterterm $\mathcal{L}$ cancels out a corresponding transformation of the data associated to the 1-form symmetries. In terms of $(A,\omega)$, this equation can be written as follows:
\begin{align}
    2\pi d\mathcal{L} &= \frac{c_-}{8}dA \wedge dA \nonumber \\
    &+ \left(\frac{k c_-}{4}+\frac{k_s}{2}\right)dA \wedge d\omega \nonumber \\
    &+ \left(\frac{c_-k^2}{8} + \frac{k_s(k_s+k)}{2}\right)d\omega \wedge d\omega .
\end{align}
We cannot directly compare these coefficients with $C,\mathscr{S}$ and $\ell$. In the usual action, we interpret $A$ as a $U(1)^f$ gauge field. However, $A$ has thus far been defined as a $U(1)_b$ gauge field. Now fermions with charge 1 under $U(1)^f$ have charge 1/2 under $U(1)_b$, so the above field theory assigns them half the correct charge. We can fix this problem by replacing $A$ with $2A$ in the expression for $\mathcal{L}$. Upon doing so, we can readily compare the above expression to the definitions of $C,\mathscr{S},\ell$. This gives
\begin{align}\label{eq:CSL}
    C &= c_- + 8 k_1 \\
    \mathscr{S} &= \frac{k c_-}{2} + k_s + 2k_2 \mod M \\
    \ell &= \frac{c_-k^2}{4} + k_s(k_s+k) + 2k_3 \mod M
\end{align}
The coefficients $k_1,k_2,k_3$ correspond to additional bosonic SPT terms that satisfy $d\nu_3 = 0 \mod 1$ and can be freely added to $\mathcal{L}$. 

\subsubsection{Crystalline equivalence principle}

We now address one final subtlety. The physical $\Z_M$ symmetry of the square lattice (for $M=4$) is a \textit{spatial} symmetry. However, the result in Eq.~\eqref{eq:CSL} applies strictly only to internal symmetries. In order to describe spatial symmetries, we need to transform $\omega_2$ according to certain rules which are formalized as the \textit{crystalline equivalence principle} for fermions (fCEP) \cite{Thorngren2018,Else2019,zhang2020realspace}.

When the point group consists only of rotations, the fCEP states that the classification of fermionic topological phases with the spatial $\Z_M$ symmetry and with integer spin fermions is identical to the classification of fermionic topological phases with an \textit{internal} $\Z_M$ symmetry but with spin-1/2 fermions, and vice versa. Thus in our formalism, we can describe a square lattice with spinless fermions by setting $k=1$, appropriate for the Hofstadter model. Setting $k=0$ would describe the square lattice with spin-1/2 fermions. Taking $k=1$, we obtain our final result for spinless fermions:

\begin{align}
    C &= c_- + 8 k_1 \\
    \mathscr{S} &= \frac{c_-}{2} + k_s + 2k_2 \mod 4 \nonumber \\
    \implies \mathscr{S} &= \frac{C}{2} \mod 1 \\
    \ell &= \frac{c_-}{4} + k_s(k_s+1) + 2k_3 \mod 4.
\end{align}
This proves our claim that $\mathscr{S} = \frac{C}{2} \mod 1$ for spinless fermions. Note that the assumption of $k=1$ (spinless fermions with respect to the spatial symmetry) is important; if we set $k=0$, $\mathscr{S}$ would be independent of $C$, and be forced to take integer values.

\subsection{Derivation of properties (2)-(3)}
Next we argue property (2). We notice that the Wen-Zee term $\frac{1}{2\pi} \epsilon^{\mu\nu\lambda} A_{\mu} \partial_{\nu} \omega_{\lambda}$ is invariant under time-reversal, which takes $\phi \rightarrow 2\pi-\phi$. $A$ and $\omega$ transform as follows:
\begin{align}
    &A_0 \rightarrow A_0, & A_i \rightarrow -A_i \\
    &\omega_0 \rightarrow -\omega_0, & \omega_i \rightarrow \omega_i.
\end{align}
The transformation of $\omega$ is obtained as follows. The angular momentum $\vec{r}\times\vec{p}$ is odd under time-reversal, and in field theory it is computed by varying the action with respect to $\omega_0$. Thus $\omega_0$ should also be odd under time reversal. Moreover, a disclination (whose disclination angle is given by the spatial integral of $\partial_1 \omega_2-\partial_2\omega_1$) is time-reversal invariant. Thus, $\omega_1,\omega_2$ should also be time-reversal invariant.
Using this, we can now check that $\epsilon^{\mu\nu\lambda} A_{\mu} \partial_{\nu} \omega_{\lambda}$ is time-reversal invariant. Thus we conclude that $\mathscr{S}$ should also be time-reversal invariant.

We will argue property (3) using Eq.~\eqref{eq:Qcube}. When $\mu$ changes sign, we have $\nu_0 \rightarrow 1-\nu_0$ in the Hofstadter butterfly. Suppose $\mathscr{S} \rightarrow \mathscr{S}'$. For convenience, consider a system on a cube, in which the total number of sites is $n_{u.c.,\text{cube}}+2$, i.e. assume there is no sublattice splitting. Then, the transformation also takes $Q_{\text{cube}} \rightarrow n_{u.c.,\text{cube}}+2 -Q_{\text{cube}}$. The original and transformed versions of Eq.~\eqref{eq:Qcube} are thus
\begin{align}
    Q_{\text{cube}} &= \nu_0 n_{u.c.,\text{cube}} + 2\mathscr{S} \\
    n_{u.c.,\text{cube}}+2 -Q_{\text{cube}} &= (1-\nu_0)n_{u.c.,\text{cube}} + 2\mathscr{S}'.
\end{align}
This gives $\mathscr{S}' = 1-\mathscr{S}$, as claimed. 
\end{document}